\documentclass[aps,prx,twocolumn,superscriptaddress,a4paper,english,longbibliography]{revtex4-2}

\usepackage[table,xcdraw]{xcolor}
\usepackage{times}
\usepackage{color}
\usepackage[colorlinks=true,urlcolor=blue,citecolor=blue]{hyperref}
\usepackage{url}
\usepackage{soul}
\usepackage{graphicx}
\usepackage{amsmath}
\usepackage{subfigure}
\usepackage{xcolor}
\usepackage{amssymb}
\usepackage{latexsym}
\usepackage{multirow}
\usepackage{hyperref}% add hypertext capabili1ties
\DeclareGraphicsExtensions{.png,.eps}
\usepackage{float}
\usepackage{ulem}
\usepackage{appendix}
\usepackage{etoolbox}
\usepackage{pifont}
\usepackage{mhchem}
\usepackage{makecell}

\usepackage{xcolor}
\usepackage[urlcolor=blue]{hyperref}      % links to citations
\hypersetup{
    colorlinks = true,                    % text and not border
    citecolor = {blue},
    linkcolor = {purple},
           }
\begin{document}	
	
\title{Intelligent diagnostic scheme for lung cancer screening with Raman spectra data by tensor network machine learning}

\author{Yu-Jia An}
\thanks{These three authors have contributed equally to this work.}
\affiliation{Department of Physics, Capital Normal University, Beijing 100048, China}

\author{Sheng-Chen Bai}
\thanks{These three authors have contributed equally to this work.}
\affiliation{Department of Physics, Capital Normal University, Beijing 100048, China}

\author{Lin Cheng}
\thanks{These three authors have contributed equally to this work.}
\affiliation{Department of Minimally Invasive Tumor Therapies Center, Beijing Hospital, National Center of Gerontology, Institute of Geriatric Medicine, Chinese Academy of Medical Sciences, Beijing, China.}
\affiliation{Savaid Medical School, University of Chinese Academy of Sciences, Beijing, China.}

\author{Xiao-Guang Li}
\affiliation{Department of Minimally Invasive Tumor Therapies Center, Beijing Hospital, National Center of Gerontology, Institute of Geriatric Medicine, Chinese Academy of Medical Sciences, Beijing, China.}

\author{Cheng-en Wang}
\email[Corresponding author. Email: ]{zhuimengzhew@163.com}
\affiliation{Department of Minimally Invasive Tumor Therapies Center, Beijing Hospital, National Center of Gerontology, Institute of Geriatric Medicine, Chinese Academy of Medical Sciences, Beijing, China.}

\author{Xiao-Dong Han}
\affiliation{Beijing Key Laboratory of Microstructure and Properties of Solids, Faculty of Materials and Manufacturing, Beijing University of Technology}

\author{Gang Su}
\affiliation{School of Physical Sciences, University of Chinese Academy of Sciences, P. O. Box 4588, Beijing 100049, China}
\affiliation{Kavli Institute for Theoretical Sciences, and CAS Center for Excellence in Topological Quantum Computation, University of Chinese Academy of Sciences, Beijing 100190, China}

\author{Shi-Ju Ran}
\email[Corresponding author. Email: ]{sjran@cnu.edu.cn}
\affiliation{Department of Physics, Capital Normal University, Beijing 100048, China}

\author{Cong Wang}
\email[Corresponding author. Email: ]{smartswang@bjut.edu.cn}
\affiliation{Beijing Key Laboratory of Microstructure and Properties of Solids, Faculty of Materials and Manufacturing, Beijing University of Technology}

\date{\today}

\begin{abstract}
Artificial intelligence (AI) has brought tremendous impacts on biomedical sciences from academic researches to clinical applications, such as in biomarkers' detection and diagnosis, optimization of treatment, and identification of new therapeutic targets in drug discovery. However, the contemporary AI technologies, particularly deep machine learning (ML), severely suffer from non-interpretability, which might uncontrollably lead to incorrect predictions. Interpretability is particularly crucial to ML for clinical diagnosis as the consumers must gain necessary sense of security and trust from firm grounds or convincing interpretations. In this work, we propose a tensor-network (TN)-ML method to reliably predict lung cancer patients and their stages via screening Raman spectra data of Volatile organic compounds (VOCs) in exhaled breath, which are generally suitable as biomarkers and are considered to be an ideal way for non-invasive lung cancer screening. The prediction of TN-ML is based on the mutual distances of the breath samples mapped to the quantum Hilbert space. Thanks to the quantum probabilistic interpretation, the certainty of the predictions can be quantitatively characterized. The accuracy of the samples with high certainty is almost 100$\%$. The incorrectly-classified samples exhibit obviously lower certainty, and thus can be decipherably identified as anomalies, which will be handled by human experts to guarantee high reliability. Our work sheds light on shifting the ``AI for biomedical sciences'' from the conventional non-interpretable ML schemes to the interpretable human-ML interactive approaches, for the purpose of high accuracy and reliability.
\end{abstract}
\maketitle

\section{Introduction}
Artificial intelligence (AI) techniques are breaking into biomedical research and health care~\cite{moore2019preparing,he2019practical,hudson2000neural}. Vast and significant potential applications of ``AI for biomedical sciences'' have been demonstrated in applying AI to image analysis in radiology, pathology, and dermatology, with diagnostic speed exceeding and accuracy paralleling medical experts~\cite{febbraro2022barriers}. Focusing on medical sciences, the deep machine learning (ML) model AlphaFold2 was developed to accurately predict protein structures at the atomic level accuracy, exhibiting superior efficiency over the human experts~\cite{jumper2021highly,tunyasuvunakool2021highly}. Yazeed Zoabi \textit{et al} proposed a ML model to identify the positive patients for SARS-CoV-2 in RT-PCR assays from eight essential features~\cite{zoabi2021machine}. Besides, ML has also made impressive progress in drug development ~\cite{reda2020machine,ekins2019exploiting}, medical image analysis~\cite{shen2017deep}, cancer research, clinical oncology~\cite{liu2017raman,bahreini2019raman,zuniga2019raman}, and etc. 

\begin{figure}[htb]
	\centering
	\includegraphics[angle=0,width=1\linewidth]{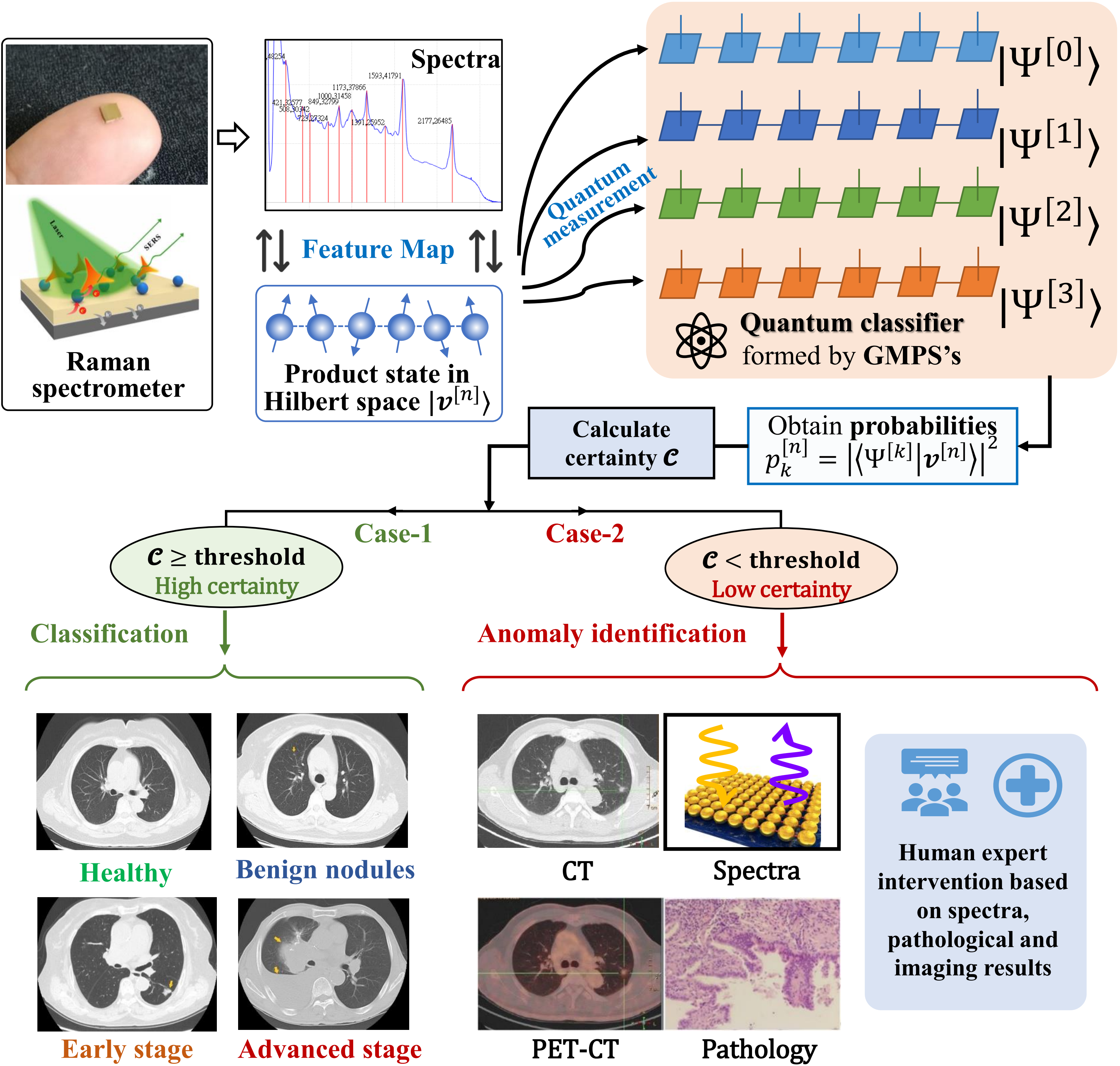}
	\caption{(Color online) The main chatflow of TN-ML for lung cancer screening. After mapping the Raman spectrometry data to quantum product states in Hilbert space, the classification probabilities (say $p_{k}^{[n]}$, the probability of the $n$-th sample belonging to the $k$-th class) can be obtained by accordingly “measuring” the generative TN states in our classifier model. The certainty of the prediction $\mathcal{C}$ is defined as the inverse of the von Neumann entropy of the distribution $\{p_{k}^{[n]}\}$ [see Eq.~(\ref{eq-certainty1})]. In the cases of high certainty, the samples are expected to be accurately classified as the healthy people group, benign nodules group, early-stage lung cancer group, or advanced lung cancer group. In the cases of low certainty (which only take up a small proportion in all the sample we tested), human-expert intervention will be introduced to make accurate judgments based on pathological, spectra, and CT imaging results.}
	\label{fig-idea}
\end{figure}

As a new paradigm, combining ML with physicians is the key route to enhance system efficiency, performance, and reliability. Especially in biomedical applications, if the diagnostic characteristics relied on decipherable diagnostic basis, human expert can support to improve identification and prediction in the performance. Otherwise, consumers may be more reluctant to utilize medical care delivered by AI providers than comparable human providers. However, the current deep ML methods, particularly deep neural networks, suffer from the notorious “non-interpretability” issue~\cite{gilpin2018explaining,zhang2018visual,carvalho2019machine}. A crucial consequence of non-interpretability is the lack of systematic access to the mathematical properties of the ML models, including the representation and generalization abilities, as well as the stability and reliability of the outcomes. The researches on methodology usually follow a ``trial-and-error'' manner that will cost significant human and computational resources. Therefore, efficient ``white-box'' ML models are strongly desired in biomedical sciences.

Recently, tensor network (TN) has been recognized as a promising candidate for interpretable quantum (or quantum-inspired) ML~\cite{SS16TNML,liu2019machine,han2018unsupervised,cheng2019tree,cheng2019tree,vieijra2022generative,sun2020generative,wang2020anomaly,vieijra2022generative}. Originating from quantum many-body physics~\cite{ran2020tensor,verstraete2008matrix,orus2019tensor,cirac2021matrix}, TN is a widely-recognized efficient numerical tool that lowers the exponential complexity of simulating quantum many-body systems to be just polynomial. Meanwhile, its quantum probabilistic nature allows to apply quantum information theories to understand the underlying mathematics in the TN methods. TN and the relevant quantum ML have been successfully applied to various scenarios that demand interpretability, such as generation~\cite{han2018unsupervised,cheng2019tree,vieijra2022generative}, feature selection~\cite{liu2021entanglement,Bai:100701}, compressed sampling~\cite{RSF+20TNCS}, anomaly detection~\cite{wang2020anomaly}, and few-shot learning~\cite{li2022non}. The few-shot scenario often appears in the application of ML to scientific problems (which is also our case) due to the inefficient data. The interpretability in these achievements is essentially based on the quantum probabilistic interpretation, which is also known as Born’s statistical interpretation.

Among the challenging and significant issues that await new AI technologies, lung cancer is globally one of the malignant tumors with the highest incidence and the leading cause of cancer death in the world (accounted for 18$\%$ of all cancer deaths in 2020)~\cite{febbraro2022barriers}. At present, the commonly used methods of lung cancer screening in clinic, such as X-ray and chest Computed Tomography (CT), are economical, but the nature of atypical lesions cannot be determined when the size of pulmonary shadow is not big enough (10-20 $mm$), which also increases the chance of radiation exposure. Low dose CT suffers and be prone to low true positive results $(\sim4\%)$ although the radiation exposure reduces to a considerable level~\cite{kaneko1996peripheral}. At present, the “gold standard” for diagnosis of lung cancer is invasive pathological examination, which is not easy to be accepted by patients with great pain, and it is difficult to meet the needs of lung cancer screening and diagnosis in the face of the population. Therefore, it is urgently needed by interpretable ML assisted with human experts to explore non-invasive, rapid, and accurate lung cancer screening methods to complement the existing lung cancer diagnostic methods and timely screen patients at early stage.

Since VOCs biomarkers from exhaled breath had been measured for the first time in 1985~\cite{gordon1985volatile}, exhalation analysis as a diagnostic method for lung cancer is a rapidly developing research field. Raman spectroscopy, as one of the analytical techniques can obtain the VOC biomarker’s scattering spectra from the incident light on surface enhanced Raman scattering substrate chip, providing the sensitive and specific biomarker molecules~\cite{yin2021efficient,smith2013raman}. Owing that Raman analysis technology is a finger-printed spectrum with the characteristic vibration and rotation peaks, recently, tremendous Raman detection have been applied to the clinical diagnosis of tumors, biological molecular structure analysis, nucleic acid, organelle and cell analysis~\cite{auner2018applications,butler2016using,schie2013label,Huang2023}, and it has shown the potential of clinical transformation application in the research of colorectal tumors, gastric cancer, breast cancer and other tumors~\cite{liu2017raman,bahreini2019raman,zuniga2019raman}. Furthermore, AI-assisted Raman detection can collect the spectra displayed by the target molecule, analyzing and extracting the characteristic spectra, so that human experts can further interpret the spectral results and check the detection results of VOCs (including aldehydes and ketones) with the help of AI, so as to confirm the clinical diagnosis of lung cancer~\cite{smith2013raman}.

In this work, we propose an ``intelligent'' screening scheme of lung cancer by TN to learn the Raman spectrometry data of VOC (see Fig.~\ref{fig-idea}). Specifically, we combine the generative TN classification (GTNC)~\cite{sun2020generative} and $t$-distributed stochastic neighboring embedding ($t$-SNE)~\cite{van2008visualizing,yang2021visualizing} schemes to reliably distinguish the healthy people and the patients with lung cancer at different stages. The predictions are obtained based on the quantum probabilistic interpretation of TN states (or called Born machines)~\cite{han2018unsupervised}. As a unique advantage of TN, its interpretability allows to define the von Neumann entropy $S$ from the quantum probabilities, which can characterize the certainty of diagnosis. 

Our numeric simulations on 130 pieces of Raman spectra of exhaled breath samples (all verified by pathological examination) show high accuracy and certainty of the prediction with clinically pathological results for most samples. Only a small proportion ($1.5\%$ for our database) of samples show low-certainty predictions. These samples are regarded as anomalies, which can be further delivered to human experts to guarantee high accuracy. The visualization by $t$-SNE explicitly demonstrates that the low-certainty samples are distributed relatively far from the clustering centers, as another explicit observation for the anomalies. Besides, the AI-assisted Raman technology is a fast, sensitive, and specific detection diagnosis, and the process of collecting respiratory samples is simple and non-invasive. Therefore, TN-ML provides an accurate, reliable, and efficient route for lung cancer screening. It is easy for patients to accept, which makes our approach time-saving and easy to be used in the physical examination for general population.

\section{CLASSIFICATION FROM RAMAN SPECTROMETRY DATA BY GENERATIVE TENSOR NETWORK CLASSIFICATION SCHEME}
    
    \subsection{Acquisition of data}
    \subsubsection{Samples}
    The breath samples were collected from healthy people and lung cancer patients in the minimally invasive Cancer Treatment Center of Beijing Hospital. This prospective and observational study was approved by the Ethics Committee of Beijing Hospital (Approval No. 2022BJYYEC-389-01), and all patients or their family members signed the informed consent. Patients with non-small cell lung cancer (NSCLC) who were admitted to Beijing Hospital from June 2022 to December 2022 and healthy subjects who were examined in the physical examination center of our hospital during the same period were selected as the research objects. A total of 130 breath samples were collected, including 25 healthy samples, 24 benign pulmonary nodules, 24 early-stage lung cancer and 57 advanced-stage lung cancer. The pathological results were deemed as the gold standard in this study, and the TNM staging of lung cancer was based on the 8th edition of the International Association for the Study of Lung Cancer~\cite{detterbeck2016iaslc}. Demographic information was collected from all subjects, including age, gender, smoking history, etc. The general information of the four groups is shown in Table~\ref{tabel-statistics}.  
    
    \begin{table*}[tbp]
    	\caption{Comparison of general data of healthy people, benign pulmonary nodules, early lung cancer, and advanced lung cancer}
    	\renewcommand\arraystretch{1.25}
    	\setlength{\tabcolsep}{3mm}{
    		\begin{tabular}{c|c|cccc}
    			\hline\hline
    			\multicolumn{2}{c|}{}       & Healthy & Benign nodules & Early stage & Advanced stage \\ \hline
    			\multicolumn{2}{c|}{$n$}    & 25    & 24    & 24    & 57                         \\ \hline
    			\multicolumn{2}{c|}{Age(y)} & $50.80\pm12.37$   & $63.33\pm8.06$    & $70.21\pm7.87$    & $69.35\pm8.16$    \\ \hline
    			\multirow{2}{*}{Gender}     & Male  & $10(40\%)$    & 14$(58.33\%)$ & 10$(41.67\%)$     & $46(80.70\%)$ \\ \cline{2-6}
    			& Female& $15(60\%)$    & 10$(41.67\%)$ & 14$(58.33\%)$     & $11(19.30\%)$     \\ \hline
    			Smoking                     & Yes       & 5 & 9             & 10                & 41            \\ \cline{2-6}
    			history                     & No    & 20            & 15            & 14                & 16                \\ \hline \hline
    	\end{tabular}}
    	\label{tabel-statistics}
    \end{table*}

    \subsubsection{Inclusion and exclusion criteria}
    Inclusion criteria: \ding{172} Patients with pulmonary space-occupying lesions confirmed by chest CT scan; \ding{173} Patients with age 18 years or older; \ding{174} Patients with no surgical contraindications and agree to receive biopsy; \ding{175} Agree to perform VOCs detection and sign the consent form.
    
    Exclusion criteria: \ding{172} Patients who had received chemotherapy or targeted therapy within 1 month before VOCs collection; \ding{173} Collected air leakage or suspected contamination; \ding{174} Patients with other tumors; \ding{175} Complicated with infection or liver diseases or kidney diseases.

\subsubsection{Data collecting methods}
    Breath samples were collected before operation treatment. The exhaled gas of each subject was passed through the self-established biochip platform based on modified atomic site as previous method~\cite{qiao2018selective, Feng2022}, and the aldehyde/ketone of the breath samples were trapped and detected by the surface-enhanced Raman scattering (SERS) chip (Fig.~\ref{fig-idea}). The sensitivity can come to ppb level and specificity can achieve over 95$\%$. All samples were analyzed on-line chip platform, and the result of Raman spectra profile will be identified by a machine learning algorithm to distinguish lung cancer patients from healthy people. Being consistent with pathological results as the gold standard, the sensitivity and specificity of machine learning-assisted Raman spectroscopy in the diagnosis of lung cancer were calculated. The AI-assisted Raman results identified as anomalies would be further checked by human experts.
 
    As a non-invasive detection technology for trace substances, Raman spectroscopy has the advantages of high sensitivity, good specificity, rapid and effective, and shows a good clinical application prospect in cancer diagnosis. In this study, we used biochip with $100 nm$ silver nanosphere array as substrate (Fig.~\ref{fig-steps}) and anchored tag molecule to capture aldehyde-ketone in breath gas (Fig.~\ref{fig-steps}), Raman scattering signal was collected by Renishaw Inc. under $1 \mu{w}/{{\mu}m^2}$ for 10 sec with ${1\mu}m$ spot diameter. To detect VOCs signal and identify four groups of people (health, benign pulmonary nodules, early-stage and advanced-stage lung cancer). It was found that compared with healthy people, patients with benign pulmonary nodules showed a similar resonance Raman peak in entire spectra. Patients with early and advanced lung cancer showed a strong resonance Raman peak at 1610-1620 and 900 ${cm}^{-1}$, the larger integrated area indicated advanced lung cancer in these characteristic Raman peak.
    
    \subsubsection{Training and testing sets}
    We randomly divide the database into the training and testing sets that contain 104 and 26 samples, respectively. The training set is used to optimize the GTNC, and the testing set will not be input to the model so that it can show the performance on the unlearnt samples. The standard deviation is estimated by 10 independent simulations with random division of the training and testing sets.
    
    \subsection{Tensor network machine learning method}
    \subsubsection{Feature map}
    As a key ingredient of our model, the generative TN state was proposed by Han \textit{et al}~\cite{han2018unsupervised} to describe the probability distribution through quantum state amplitudes. Sun \textit{et al} proposed GTNC by utilizing multiple generative TN states to implement classification tasks~\cite{sun2020generative}. Considering $N$ samples $\{\boldsymbol{x}^{[n]}\}$ $(n=1,\ldots,N)$ where each sample contains $M$ features $\boldsymbol{x}^{[n]} = \{x_m^{[n]}\}$ ($m=1,\ldots,M$), the first step is to normalize the data so that $x_m^{[n]}\in[0,1]$. Then we map each sample $\{\boldsymbol{x}^{[n]}\}$ to a quantum product state of $M$ sites~\cite{SS16TNML} as
    \begin{eqnarray}
        |{\varphi}^{[n]}\rangle = \prod_{m=1}^M \sum_{s=1}^{d} 
        {v}^{[n,m]}_{s} |{s}_m\rangle,
        \label{eq-Sample direct product state}
    \end{eqnarray}
where the state vectors $\{|{s}_m\rangle\}$ $(s_m=0,\ldots,d-1)$ define an orthonormal basis for the $m$-th site, and the hyperparameter $d$ controls the degrees of freedom of each site. For $d=2$, each site represents a spin-1/2 or qubit. The coefficients satisfy
    \begin{eqnarray}
        v^{[n,m]}_{s} = \sqrt{\binom{d-1}{s-1}} \cos \left( \frac{\theta \pi}{2} x^{[n]}_m \right)^{d-s} \sin \left( \frac{\theta \pi}{2} x^{[n]}_m \right)^{s-1}.
        \label{eq-featuremap}
    \end{eqnarray}
    where the hyperparameter $\theta$ controls the maximal angle of each single-site state $\sum_{s=1}^{d} {v}^{[n,m]}_{s} |{s}_m\rangle$ in the Bloch sphere. Both hyperparameters affect the strength of quantum fluctuations in the TN model, and their values can be properly determined according to the training set. In this work, we take $d = 6$ and $\theta = 1.5$.
    
    \subsubsection{Training GTNC for classification}
    
    A GTNC is formed by multiple generative TN states. Here, we take each state in the form of matrix product state that is defined by $M$ local tensors $\{\boldsymbol{A}^{[M]}\}$ $(m=1,\ldots,M)$ as
    \begin{eqnarray}
        |\Psi\rangle =  \sum_{s_{1} \ldots s_{M}=0}^{d-1} \sum_{a_{1} \ldots a_{M-1}=0}^{\chi-1} A^{[1]}_{s_{1} a_{1}} A^{[2]}_{s_{2} a_{1}a_{2}} \ldots \nonumber \\ A^{[M-1]}_{s_{M-1} a_{M-2}a_{M-1}} A^{[M]}_{s_{M} a_{M-1}} \prod_{m=1}^M |{s}_m\rangle.
        \label{eq-MPS}
    \end{eqnarray}
    The indexes $\{s_m\}$ $(m=1,\ldots,M)$ are called physical indexes with $\dim{\left(s_m\right)}=d$. The indexes $\{\alpha_m\}$ are called virtual indexes, and their dimension (denoted as $\chi$) is a hyperparameter that determines complexity of the TN state and the maximal entanglement entropy the TN state can carry. 
    
    Considering, e.g., the $n$-th sample $\boldsymbol{x}^{\left[n\right]}$, the probability of generating this sample from a given TN state $|\Psi\rangle$ satisfies
    \begin{eqnarray}
        &&P\left(\boldsymbol{x}^{\left[n\right]}\right) = \left| \langle \varphi^{\left[n\right]}|\mathrm{\Psi} \rangle \right|^{2} \nonumber \\ &&= \left| \sum_{s_{1} \ldots s_{M}=0}^{d-1} \sum_{a_{1} \ldots a_{M-1}=0}^{\chi-1} A^{[1]}_{s_{1} a_{1}} \ldots A^{[M]}_{s_{M} a_{M-1}} \prod_{m=1}^M {v}^{[n,m]}_{s} \right|^{2}.\nonumber \\ 
        \label{eq-classical probability}
    \end{eqnarray}
    This is also the probability of projecting $|\Psi\rangle$ to $| \varphi^{\left[n\right]} \rangle$ by the means of quantum measurement. Considering a classification problem with $K$ classes, one constructs a GTNC model formed by $K$ generative TN states $|\Psi^{[k]}\rangle$ and uses Eq.~(\ref{eq-classical probability}) to obtain the probability of $\boldsymbol{x}^{[n]}$ belonging to the $k$-th class as
    \begin{eqnarray}
        P_k^{[n]} = \left| \langle \varphi^{\left[n\right]}|\mathrm{\Psi}^{[k]} \rangle \right|^{2}.
        \label{eq-pk}
    \end{eqnarray}
    The prediction of the classification ${\widetilde{k}}$ for $x^{[n]}$ is given by
    \begin{eqnarray}
        {\widetilde{k}} = \mathop{\mathrm{argmax}}\limits_{k} {P_k^{[n]}}
        \label{eq-argmax}
    \end{eqnarray}
    
    To optimize the tensors in the TN, we define the loss function of the training set as the negative logarithmic likelihood
    \begin{eqnarray}
        L = -\frac{1}{N} \sum_{n{\in} \text{training set}} \ln{P_{\tilde{k}}^{[n]}},
        \label{eq-NLL}
    \end{eqnarray}
    where $\tilde{k}$ denotes the ground-truth classification of the $n$-th training sample. In the extreme case with $L=0$, all training samples will be classified correctly and $P_{\tilde{k}}^{[n]}$ gives the one-hot distribution. Thus, the optimization task is to minimize $L$ by updating the tensors $\{\boldsymbol{A}^{[m]}\}$ using gradient descent as
    \begin{eqnarray}
        \boldsymbol{A}^{\left[m\right]}\gets{\boldsymbol{A}}^{\left[m\right]}-\eta\frac{\partial L}{\partial{\boldsymbol{A}}^{\left[m\right]}}
        \label{eq-grad}
    \end{eqnarray}
    The update is implemented using a sweep scheme inspired by the density matrix renormalization group method~\cite{SS16TNML,W92DMRG}. With a low $L$, the model is expected to accurately predict the classifications of not only the training samples but also the testing samples that are not used to optimize the TN states.

    \section{RESULTS AND DISCUSSIONS}
    
    \subsection{Accurate classification with certainty}
    \begin{figure}[tbp]
    	\centering
    	\includegraphics[angle=0,width=1\linewidth]{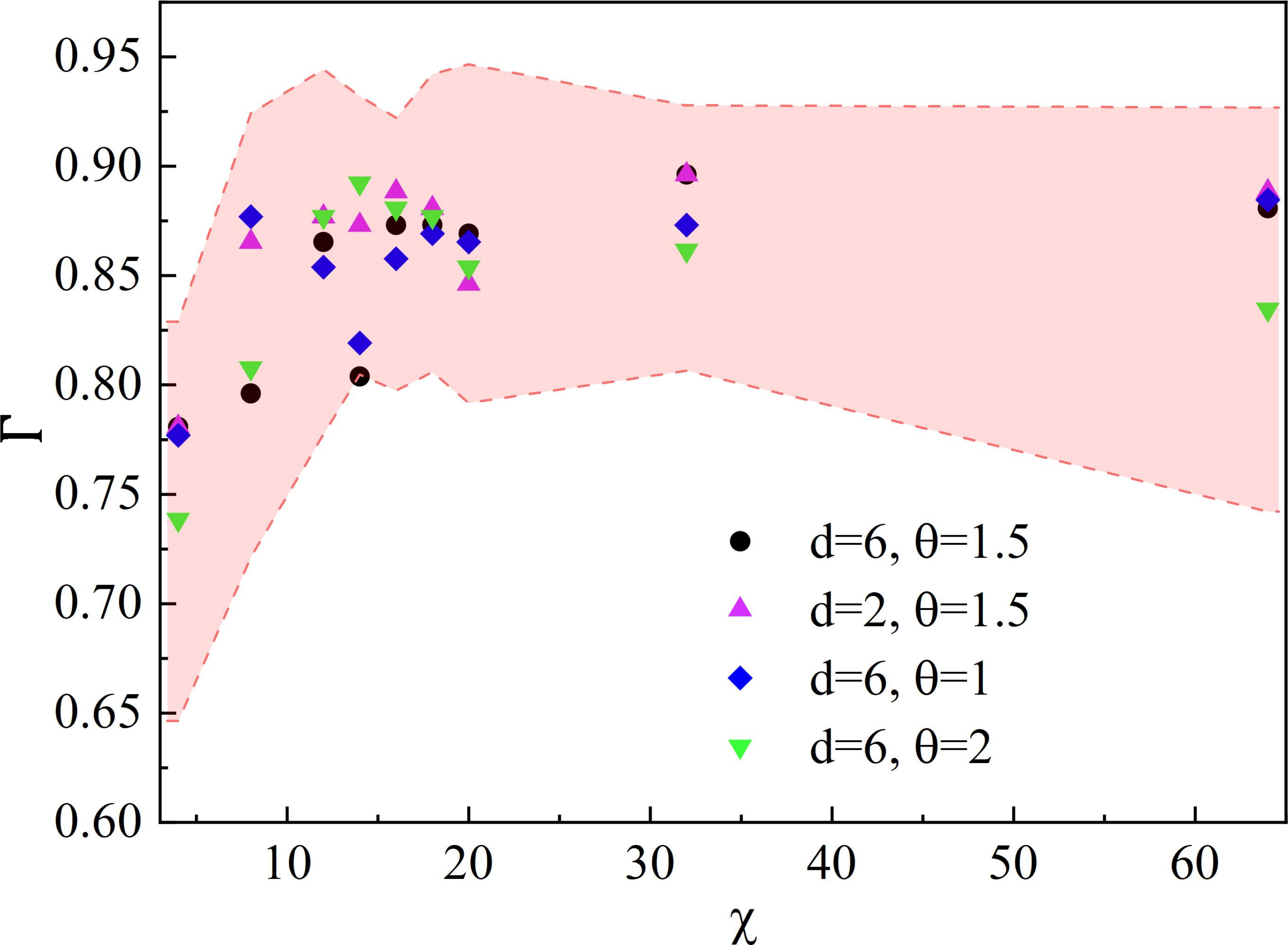}
    	\caption{(Color online) The testing accuracy $\mathrm{\Gamma}$ against the virtual dimension $\chi$ of GTNC with different $d$ and $\theta$. Each point represents the average over ten independent simulations by randomly defining the training and testing sets. The red shadow shows the standard deviation of all results by the same $\chi$.}
    	\label{fig-hyperparameters}
    \end{figure}
    
    The accuracy for the training set of GTNC is about 99.4$\%$, which characterizes the representation ability, i.e., how well the GTNC can classify the samples it has learnt. The accuracy for the testing set characterizes generalization ability of the GTNC for handling the unlearnt samples. Fig.~\ref{fig-hyperparameters} shows the testing accuracy by varying the virtual bond dimension $\chi$ (i.e., the parameter complexity of GTNC). The red shadow shows the standard deviation (\textit{std}) of all results by the same $\chi$. For about $\chi>10$, the testing accuracy GTNC is stably around $80\% \sim 90\%$. 
    
\begin{figure*}[tbp]
       \centering
       \includegraphics[angle=0,width=0.6\linewidth]{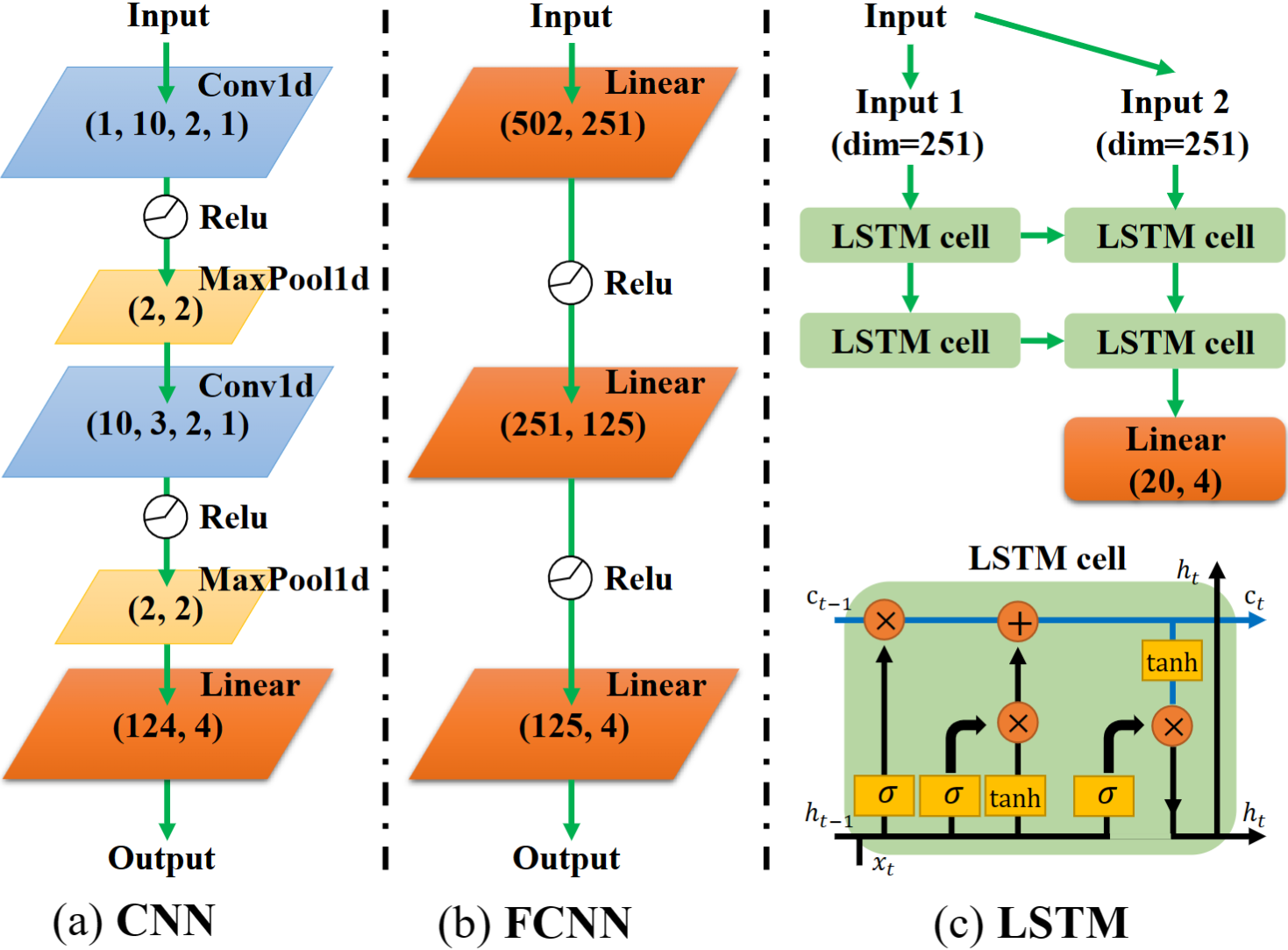}
       \caption{(Color online)  The structure and hyper-parameters of the NN models used in Table \ref{tabel-classification accuracy}, which are (a) one-dimensional convolutional NN (CNN), (b) fully-connected NN (FCNN), and (c) long-short term memory (LSTM).}
       \label{fig-NN}
    \end{figure*}

    In Table~\ref{tabel-classification accuracy}, we compare the training and testing accuracies of GTNC with several widely-recognized NN models, which are one-dimensional convolutional NN (CNN), long-short term memory (LSTM), and fully-connected NN (FCNN)~\cite{sainath2015convolutional}. The details (structure and hyper-parameters) of these models are shown in Fig.~\ref{fig-NN}. Similar accuracies will be obtained by adjusting these models.
    
    \begin{table}
        \caption{Training, testing accuracies and the standard deviation (\textit{std}) obtained by GTNC $(d = 6, \theta = 1.5)$, convolutional neural network (CNN), long short-term memory (LSTM), and fully-connected neural network (FCNN). The hyper-parameters of these ML models have been properly adjusted.}
        \begin{tabular*}{8.5cm}{@{\extracolsep{\fill}}r|cccc}
        \hline \hline
         & GTNC   & CNN    & LSTM   & FCNN   \\ \hline
        Training accuracy & 0.9942 & 0.9394 & 1      & 0.9933 \\ \hline 
        \textit{std}      & 0.0907 & 0.1783 & 0      & 0.0091 \\ \hline
        Testing accuracy  & 0.8731 & 0.8577 & 0.8692 & 0.8846 \\ \hline
        \textit{std}     & 0.0907 & 0.1602 & 0.0579 & 0.0544 \\ \hline \hline
        \end{tabular*}
        \label{tabel-classification accuracy}
    \end{table}

    The unique advantage of GTNC over NN models is the quantum probabilistic interpretability. We define the certainty of prediction by the von Neumann entropy $S$ of the $K$ quantum probabilities $P_k\left({x}^{\left[n\right]}\right)$ [Eq.~(\ref{eq-classical probability})] with
    \begin{eqnarray}
        S=\frac{1}{Z^{\left[n\right]}}\sum_{k=1}^{K}\left[P_k\left({x}^{\left[n\right]}\right)\ln{Z^{\left[n\right]}} - P_k\left({x}^{\left[n\right]}\right)\ln{P_k\left({x}^{\left[n\right]}\right)}\right],\nonumber \\
        \label{eq-certainty}
    \end{eqnarray}
with $Z^{[n]} = \sum_{k=1}^{K} P_k({x}^{[n]})$ the normalization factor. We define the certainty as the negative logarithmic entropy 
    \begin{eqnarray}
        \mathcal{C}=- \ln S.
        \label{eq-certainty1}
    \end{eqnarray}
In Table~\ref{tabel-certainty}, we show in an ascending order the $10$ smallest certainties in the testing set. The only two incorrectly classified samples give the two smallest certainties with $\mathcal{C}=0.029$ and $0.591$, and the rest are all classified correctly by GTNC with $\mathcal{C}>1$. Our results suggest that high accuracy can be obtained by dividing the samples into ``certain'' and ``uncertain'' groups. Most of the predictions show high certainty (say $\mathcal{C}>1$), which can be classified with high accuracy ($100\%$ in our database). Correspondingly, the incorrectly-classified samples are mostly likely in the uncertain group.

\begin{table*}[tpb]
 \caption{The 10 smallest certainties (first row) in the testing set, the corresponding ground-truth classifications of these samples (second row), and their predictions by GTNC (third row). The first two columns show the only two incorrectly classified testing samples.}
 \renewcommand\arraystretch{1.25}
 \begin{tabular*}{16cm}{@{\extracolsep{\fill}}c|cccccccccc}
 \hline \hline                        
    \textbf{Certainty} $\mathcal{C}$ & 0.029       & 0.591  & 1.119  & 1.162  & 2.145  & 3.081  & 3.531  & 3.640  & 4.313  & 5.638  \\ \hline
 \textbf{Classification (label)}  & 3  & 3  & 0  & 3  & 2  & 3  & 3  & 0  & 2  & 3             \\ \hline
    \textbf{Prediction}                                    & {\color[HTML]{FF0000} \textbf{0}} & {\color[HTML]{FF0000} \textbf{0}} & {\color[HTML]{00B050} \textbf{0}} & {\color[HTML]{00B050} \textbf{3}} & {\color[HTML]{00B050} \textbf{2}} & {\color[HTML]{00B050} \textbf{3}} & {\color[HTML]{00B050} \textbf{3}} & {\color[HTML]{00B050} \textbf{0}} & {\color[HTML]{00B050} \textbf{2}} & {\color[HTML]{00B050} \textbf{3}} \\ \hline \hline
    \end{tabular*}
    \label{tabel-certainty}
\end{table*}

Based on the above observations, we propose the following TN-ML diagnostic scheme with high accuracy and reliability. Given a sample (denoted by $\boldsymbol{y}$) that is to be classified, the first step is to calculate the probabilities $P_k(\boldsymbol{y})$ [Eq.~(\ref{eq-pk})] using a trained GTNC model. The second step is to calculate the certainty of this sample $\mathcal{C}$ [Eq.~(\ref{eq-certainty1})]. If the certainty is larger than a preset threshold (say $\mathcal{C}=0.85$, the average of second and third smallest certainties in Table~\ref{tabel-certainty}), one calculates its classification [Eq.~(\ref{eq-argmax})], which will be highly accurate. If the certainty is smaller than the threshold, one sends an alert for re-examination or deliver the sample to the hands of human experts. We call such samples as anomalies. This TN-ML diagnostic scheme is expected to accurately handle most of the samples efficiently by ML, and only leaves a very small proportion of samples for further treatments by human experts.

\subsection{Visualization}
\begin{figure}[tbp]
   \centering
   \includegraphics[angle=0,width=1\linewidth]{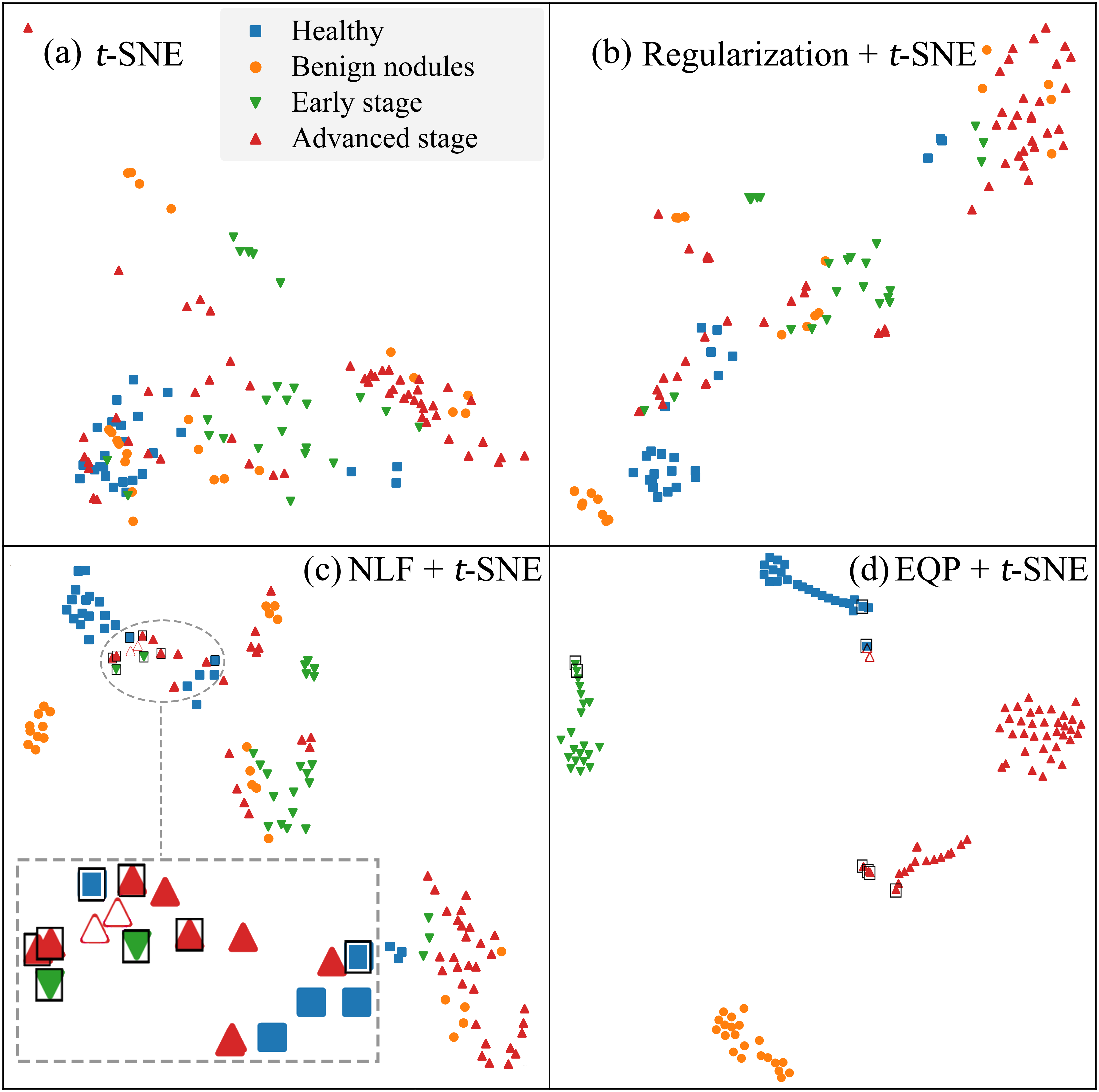}
   \caption{(Color online) The visualization of data distribution by $t$-SNE with different kernel functions. In (a) and (b), we adopt the Euclidean distance [Eq.~(\ref{eq-Euclidean})] with and without regularizing values of features to be $0 \leq x_m^{[n]} \leq 1$, respectively. In (c) and (d), we use the NLF [Eq.~(\ref{eq-nlf})] and EQP [Eq.~(\ref{eq-eqp})]calculated from by GTNC for dimensionality reduction. In the insets of (c), we show the zoom-in of the samples with low certainty. The hollow symbols indicate the anomalies with low certainty and incorrect classification, and those in black box indicate the samples with low certainty but correct classification.}
   \label{fig-tsne}
\end{figure}

    We further analyze the anomalies by visualizing the distribution of samples in an effective feature space by different non-linear dimensionality reduction mappings. Visualization will provide intuitional evidence that can be easily understood by both the physicians and patients. Specifically, we use $t$-SNE with different kernel functions. In Fig.~\ref{fig-tsne} (a) and (b), we define the kernel function by the Euclidean distance 
\begin{eqnarray}
        D_{n_1n_2}^{\text{E}}=\left|{\boldsymbol{x}}^{\left[n_1\right]}-{\boldsymbol{x}}^{\left[n_2\right]}\right|,
        \label{eq-Euclidean}
    \end{eqnarray}
with and without regularizing values of features to be $0 \leq x_m^{[n]} \leq 1$. Regularization can slightly improve the clustering, meaning the samples in a same class tend to gather to each other in the effective feature space. 

    In Fig.~\ref{fig-tsne}(c), we adopt the negative logarithmic fidelity (NLF) as the measure of distance defined as
    \begin{eqnarray}
        D_{n_1n_2}^{\text{NLF}}=-ln\left|\left\langle\varphi^{\left[n_1\right]}\right.|\left.\varphi^{\left[n_2\right]}\right\rangle\right|.
        \label{eq-nlf}
    \end{eqnarray}
Such a distance has been used in the optimization of TN and quantum circuit models~\cite{R20MPSencode,yang2021visualizing,li2022non}, and much improves the clustering in our scenario. The inset shows the zoom-in of distributions of the samples with low certainty. The hollow symbols indicate the anomalies with low certainty and incorrect classification, and those in black box indicate the samples with low certainty but correct classification. Obviously, the ``uncertain'' samples, particularly the anomalies, are located at the “edges” of the clusters.
    
In Fig.~\ref{fig-tsne}(d), we use the Euclidean distance between quantum probabilities $\boldsymbol{P}^{[n]}$ (EQP)~\cite{li2022non}, and achieve the clearest clustering. The distance of two samples is then defined as
\begin{eqnarray}
  D_{n_1n_2}^{\text{EQP}} = \sum_k \left|\frac{P_k(\boldsymbol x^{[n_1]})}{Z^{[n_1]}} - \frac{P_k(\boldsymbol x^{[n_2]})}{Z^{[n_2]}} \right|^2 .
   \label{eq-eqp}
\end{eqnarray}
The two anomalies are at the edge of the cluster for the class of healthy people. This also visually shows that these two anomalies are relatively close to the cluster of the healthy samples, which interprets the misclassifications (see Table~\ref{tabel-certainty}).
    
\subsection{Discussion with human-expert intervention}
\begin{figure}[tbp]
  \centering
   \includegraphics[angle=0,width=1\linewidth]{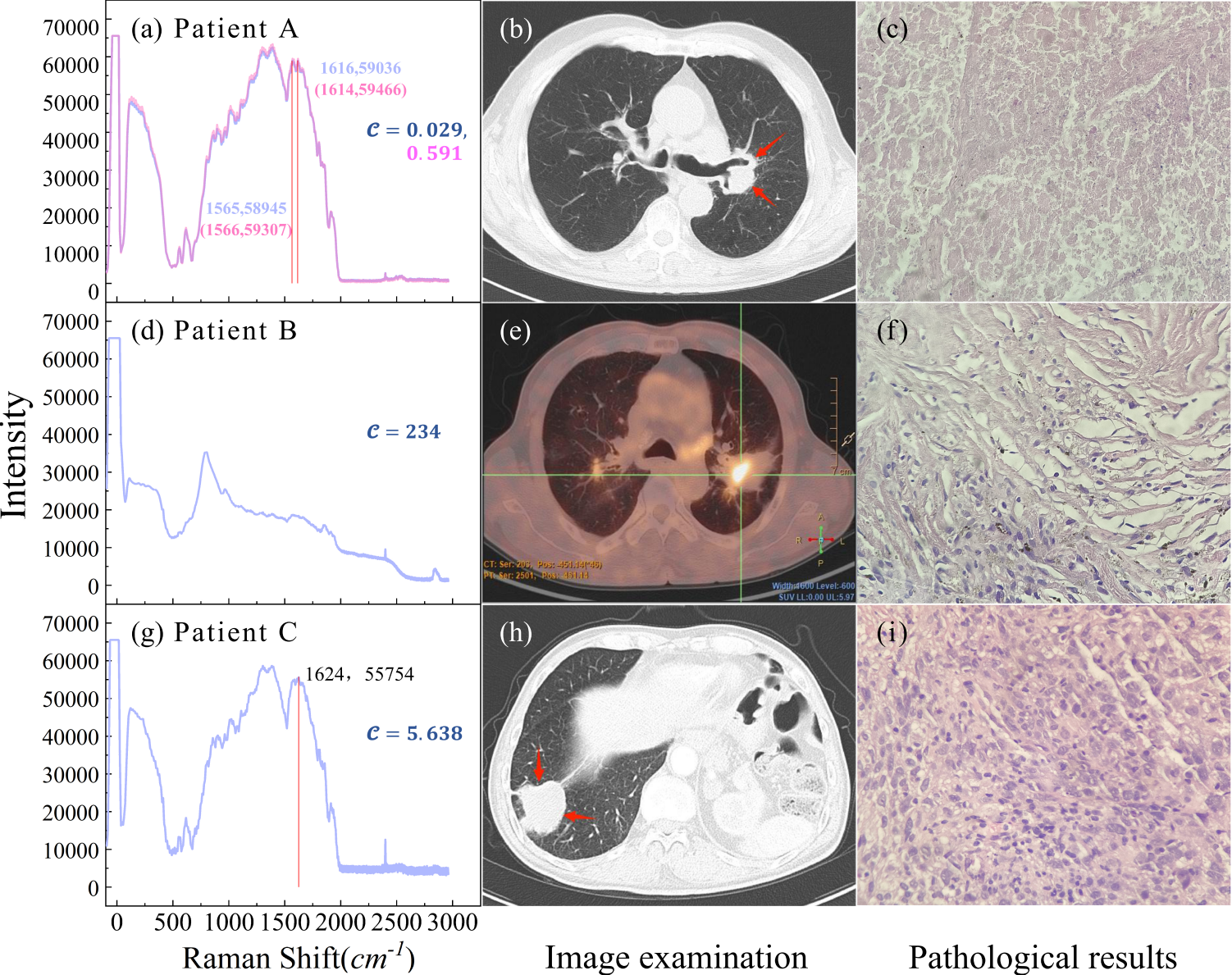}
   \caption{(Color online) The Raman spectra, image examinations, and pathological results. In (a) we show the Raman spectra from two tests on patient A after proper treatment (in blue and purple, respectively), which give the two lowest certainty by TN-ML (see the first two columns in Table \ref{tabel-certainty}). The image examination, and pathological result of patient A are shown in (b) and (c). This patient was diagnosed to be in the advanced stage, but obtained great therapeutic effects after treatment. The second [(d)-(f)] row shows the results of patient B who is suspected to have cancer by PET-CT before pathological, and finally diagnosed to be in the benign nodules stage. The third row [(g)-(i)] shows the results of patient C in the advanced stage. The red arrows indicate the serum tumor markers in the first and third patients.}
   \label{fig-laman}
\end{figure}

\begin{figure}[tbp]
   \centering
   \includegraphics[angle=0,width=1\linewidth]{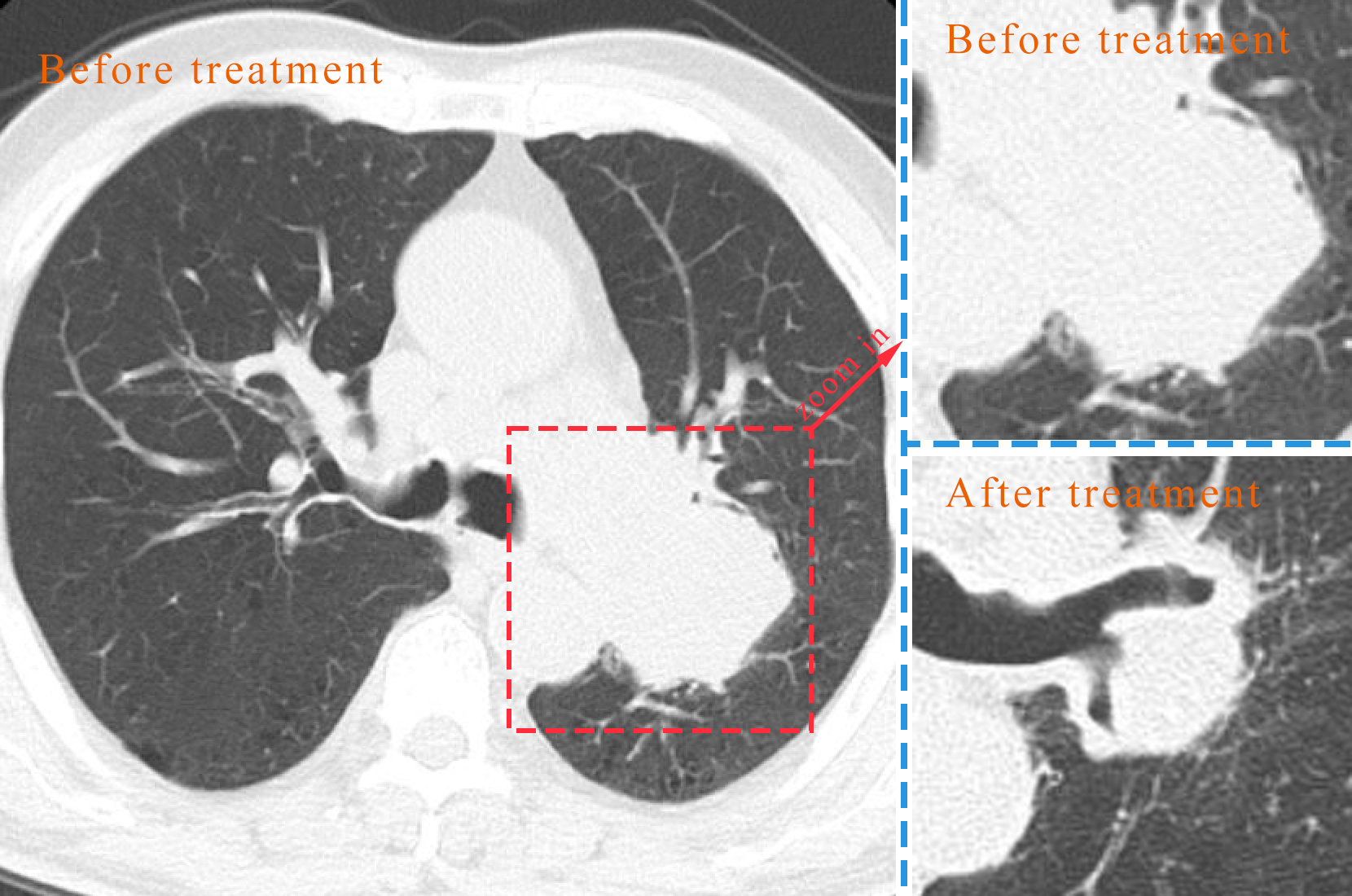}
   \caption{(Color online) The left side shows the CT image of patient A (advanced lung cancer) before treatment. The right side demonstrates the changes in the patient's lung tumor before and after treatment.}
   \label{appendix fig.4}
\end{figure}

 Our TN-ML Raman analysis results are over $98.5\%$ consistent with the pathological results. When the analysis result is abnormal, human experts could be intervened based on Raman characteristic spectra and other pathological results. In this work, according to our TN-ML results, there are only two abnormal data with low certainty [see the Raman spectra, image examination, and pathological results in Fig.~\ref{fig-laman}(a)-(c)]. These two samples are from a same patient (denoted as A) during medical treatment, who was previously diagnosed to be in the advanced stage but obtained great therapeutic effects by treatment. Hence, human expert can consider this patient as advanced lung cancer according to previous CT (Fig.~\ref{appendix fig.4}) and pathological results [Fig.~\ref{fig-laman}(c)]. However, during the medical treatment, it is a question that what stage is he/she at this time. The AI-assisted Raman spectra with characterized peaks in Fig.~\ref{fig-laman}(a) can give a diagnosis that the peaks of $1610$-$1620$ $cm^{-1}$ are vital but weak, which are specific peaks for the targeted aldehyde’s reaction with \ce{-NH2} probe on chips. Although it is not obvious for human expert in Fig.~\ref{fig-laman} (a), patient A is considered to have lung cancer. When the clinical medical data of this patient [CT in Fig.~\ref{fig-laman}(b) and serum tumor marker in Fig.~\ref{fig-laman}(c)] are checked, accurate clinical diagnoses can be made. This explains the TN misclassification of these two samples to the healthy group and the low certainty, as well as the visualization results in Fig.~\ref{fig-tsne}(d) (see the hollow symbols). In this case our TN-ML helps human expert to notice the suspected cases and increase the certainty of diagnoses. From the other side, the human experts will only need to handle these two anomalies so that the accuracy can be improved to $100\%$. This workflow can be continuously used to analyze big data in further investigations.

The second row of Fig.~\ref{fig-laman} shows the results of a patient (denoted as B) diagnosed to be in the benign nodules stage. Based on the Raman result [Fig.~\ref{fig-laman}(d)] that shows weaker intensity of the characterized peaks, the patient is suspected of having benign lesions. The PET-CT result (Fig.~\ref{fig-laman}(e)) indicates the advanced lung cancer, but the pathological results (Fig.~\ref{fig-laman}(f)) show no tumor cells. Considering the pathological results are the gold standard for clinical diagnosis, this sample is labeled as benign nodules. Our TN-ML classifies this sample correctly with high certainty ($\mathcal{C} = 234 \gg 0.85$), which is consistent with pathological results. In this case our TN-ML helps human expert to identify the possible CT image error to increase the diagnosis certainty. 

The third row of Fig.~\ref{fig-laman} shows the results from patient C with a typical advanced lung cancer. The Raman spectrum clearly shows a $1600$ $cm^{-1}$ peak and more intensive peaks around $990$ $cm^{-1}$. These could be ascribed to surface active sites aggregation on the sensor chip, only occurring under high concentrations of volatile organic compounds. Image examination and pathological results are consistent with the Raman spectra. In this case, our TN-ML manages to make correct prediction and the certainty is much larger than the threshold ($\mathcal{C}=5.638 \gg 0.85$).

\begin{table}[tbp]
    \caption{The serum tumor markers of patients A, B, and C, including Neuron specific enolase (NSE), cytokeratin fragment 21-1 (CYFRA21-1), carcinoembryonic antigen (CEA), carbohydrate antigen 125 (CA125), Cancer Antigen 15-3 (CA15-3), and cancer antigen 19-9 (CA19-9). For patient A, all indexes turned to be normal after treatment. For patient B, the benign nodules cannot be distinguished from these indexes as they are all normal. For patient C, these indexes clearly imply the advanced lung cancer.}
    \renewcommand\arraystretch{1.25}
    \begin{tabular}{c|c|c|c|c}
    \hline \hline
    \textbf{Patient}   & \multicolumn{2}{c|}{A}   & B          & C    \\  \hline
    \textbf{Stage}     & \multicolumn{2}{c|}{Advanced stage}   & Benign nodules  & Advanced stage    \\  \hline
    \textbf{Treatment}   & before      & after      & before     & before \\ \hline
    \textbf{NSE (ng/ml)}     & {\color[HTML]{FF0000} \textbf{$34.4\uparrow$}}    & 11    & 13.8   & 14.4  \\ \hline
    \textbf{\makecell{CYFRA \\ 21-1 (ng/ml)}}   & {\color[HTML]{FF0000} \textbf{$37.9\uparrow$}}    & 2.72  & 2.53   & {\color[HTML]{FF0000} \textbf{$3.35\uparrow$}}  \\ \hline
    \textbf{CEA (ng/ml)}     & 3.3   & 2.6   & 3.7   & {\color[HTML]{FF0000} \textbf{$11.2\uparrow$}}    \\ \hline
    \textbf{CA125 (U/ml)}   & {\color[HTML]{FF0000} \textbf{$41.8\uparrow$}}    & 16.9  & 16.2   & {\color[HTML]{FF0000} \textbf{$44.8\uparrow$}}    \\ \hline
    \textbf{CA15-3 (U/ml)}  & 6.4   & 5.3   & 9.5   & 12.5  \\ \hline
    \textbf{CA19-9 (U/ml)}  & $<2$   & 5.5  & 13    & 3.6   \\ \hline \hline
    \end{tabular}
    \label{table-details}
\end{table}

Additional examination indexes on several serum tumor markers of patients A, B, and C are shown in Table \ref{table-details} (serum tumor markers) and Fig.~\ref{appendix fig.4} (CT images). These indexes of A turned to be normal after treatment, which again explains why TN-ML classifies the corresponding samples (the two anomalies with low certainty) to the healthy group. For patient B, one cannot tell the benign nodules from these indexes as they all appear to be normal. For C, clear signals for the advanced lung cancer are demonstrated by these indexes. Except for the image and pathological examinations, our TN-ML with Raman spectral data can make accurate and reliable diagnoses itself, which can be potentially applied in population screening in primary hospital or remote region. 

\section{Summary}

\begin{figure*}[tbp]
  \centering
   \includegraphics[angle=0,width=1\linewidth]{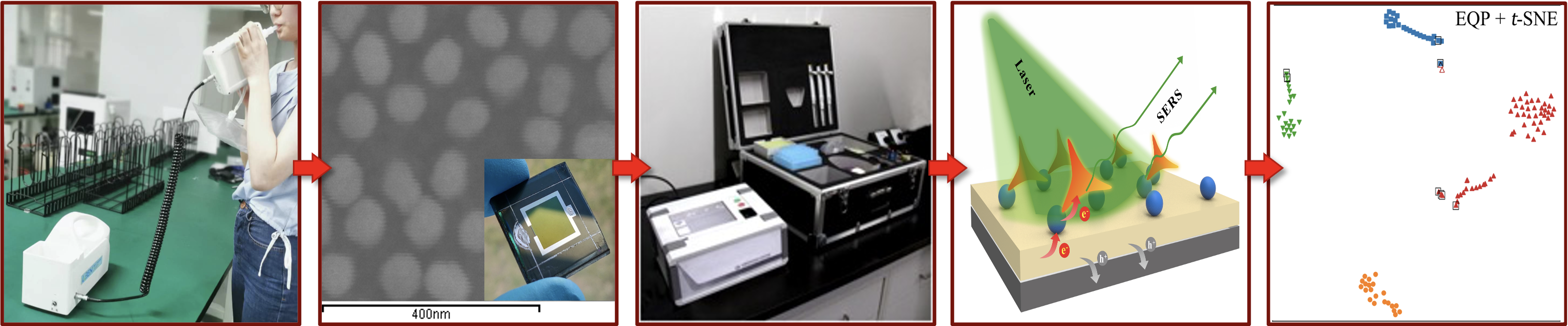}
   \caption{(Color online) The main steps of our ML-assisted diagnostic scheme: capture and detection on chip, Raman spectra collection, and ML-assisted simulations. The details of the last step are illustrated in Fig.~\ref{fig-idea}.}
   \label{fig-steps}
\end{figure*}

We here proposed a AI-assisted diagnostic scheme for lung cancer screening (see the main steps in Fig.~\ref{fig-steps}). At present, the commonly used clinical diagnostic methods for lung cancer include blood tumor markers, chest X-ray, chest CT, fiberoptic bronchoscopy, percutaneous lung biopsy, etc. Although they have shown good clinical utility in the diagnosis of lung cancer, there are more or less problems such as poor sensitivity, low accuracy, high cost, and invasiveness. Therefore, it is of great clinical significance to develop a sensitive, specific, non-invasive, simple, reliable diagnostic technique. 
 
 ML-assisted techniques provide a broad platform for accuracy and efficiency improvement. Decipherable method has additional benefit for professional intervention, which can further enhance its performance and solve the vital issue of “non-interpretability” AI in medical applications. The composition of exhaled gas can reflect the metabolism of corresponding tissues and cells. The metabolic process and products of cancerous cells are different from those of normal cells, which is helpful to detect diseases by detecting the changes of exhaled gas composition. Up to date, certain VOCs biomarkers related to lung cancer, which is helpful for the diagnosis~\cite{cirac2021matrix,han2018unsupervised,cheng2019tree}. The Raman-spectra driven non-invasive and sensitive technique has great potential to detect metabolic disease or cancer cell metabolome. More sensitive and specific active probe molecules pre-grafted on Raman chip are well-noted to improve gaseous aldehyde or ketone molecules captured and achieve detection at lower than parts per billion levels. Therefore, TN-ML assisted Raman detection combined with human expert intervention technique makes it a great potential for rapid, simple, economical, and noninvasive identification of cancer and drug treatment monitoring. 
 
In our work, the prediction of TN-ML is made by quantitatively characterizing the mutual distances of the breath samples after mapping to the quantum Hilbert space. The sensitivity and certainty of the predictions can be characterized by utilizing the “quantum” probabilistic interpretation of the TN-ML model, where the decipherable Raman spectra data with low certainty can be identified as anomalies that will be handled by human experts. Furthermore, human expert intervention confirms the meaningful spectra with low certainty for further improvement of accuracy. Our work shows that Raman technology assisted by the ``white-box'' ML can collectively improve the accuracy and efficiency of diagnosis and expand the range of indications.

\section*{Acknowledgment} 
SCB is grateful to Yi-Cheng Tang for helpful discussions. This work is supported by NSFC (Grant No. 12004266, No. 11834014 and No. 12104023), Foundation of Beijing Education Committees (Grant No. KM202010028013), the Strategic Priority Research Program of the Chinese Academy of Sciences (Grant No. XDB28000000), the National Key R\&D Program of China (Grant No. 2018YFA0305800), the key research project of Academy for Multidisciplinary Studies, Capital Normal University.

\section*{Competing Interests}
The authors declare no competing financial or non-financial interests.

\section*{Data Availability}
The authors will provide the codes and data under reasonable requests. 

\section*{Author Contributions}
YJA, SCB, GS, and SJR contributed to the tensor network and neural network machine learning. LC and CEW contributed to the breath samples and patients' results collection. CEW, XDH, SJR and CW conceived the idea and designed the study. All authors contributed to preparing the manuscript. 

% \section{General information of database}

%\section{Details of neural networks}

%To be added.

%\section{Image examinations of patient A before and after treatment}

%\begin{figure}[tbp]
%\centering
%\includegraphics[angle=0,width=1\linewidth]{fig5.png}
%\caption{(Color online) The Raman spectra, image examinations, and pathological results. In (a) we show the Raman spectra from two tests on patient A after proper treatment (in blue and purple, respectively), which give the two lowest certainty by TN-ML (see the first two columns in Table \ref{tabel-certainty}). The image examination, and pathological result of patient A are shown in (b) and (c). This patient was diagnosed to be in the advanced stage, but obtained great therapeutic effects after treatment. The second [(d)-(f)] row shows the results of patient B who is suspected to have cancer by PET-CT before pathological, and finally diagnosed to be in the benign nodules stage. The third row [(g)-(i)] shows the results of patient C in the advanced stage. The red arrows indicate the serum tumor markers in the first and third patients.}
%		\label{fig-tmp}
%\end{figure}

\normalem

% \bibliography{bibi}

\begin{thebibliography}{47}%
\makeatletter
\providecommand \@ifxundefined [1]{%
 \@ifx{#1\undefined}
}%
\providecommand \@ifnum [1]{%
 \ifnum #1\expandafter \@firstoftwo
 \else \expandafter \@secondoftwo
 \fi
}%
\providecommand \@ifx [1]{%
 \ifx #1\expandafter \@firstoftwo
 \else \expandafter \@secondoftwo
 \fi
}%
\providecommand \natexlab [1]{#1}%
\providecommand \enquote  [1]{``#1''}%
\providecommand \bibnamefont  [1]{#1}%
\providecommand \bibfnamefont [1]{#1}%
\providecommand \citenamefont [1]{#1}%
\providecommand \href@noop [0]{\@secondoftwo}%
\providecommand \href [0]{\begingroup \@sanitize@url \@href}%
\providecommand \@href[1]{\@@startlink{#1}\@@href}%
\providecommand \@@href[1]{\endgroup#1\@@endlink}%
\providecommand \@sanitize@url [0]{\catcode `\\12\catcode `\$12\catcode
  `\&12\catcode `\#12\catcode `\^12\catcode `\_12\catcode `\%12\relax}%
\providecommand \@@startlink[1]{}%
\providecommand \@@endlink[0]{}%
\providecommand \url  [0]{\begingroup\@sanitize@url \@url }%
\providecommand \@url [1]{\endgroup\@href {#1}{\urlprefix }}%
\providecommand \urlprefix  [0]{URL }%
\providecommand \Eprint [0]{\href }%
\providecommand \doibase [0]{https://doi.org/}%
\providecommand \selectlanguage [0]{\@gobble}%
\providecommand \bibinfo  [0]{\@secondoftwo}%
\providecommand \bibfield  [0]{\@secondoftwo}%
\providecommand \translation [1]{[#1]}%
\providecommand \BibitemOpen [0]{}%
\providecommand \bibitemStop [0]{}%
\providecommand \bibitemNoStop [0]{.\EOS\space}%
\providecommand \EOS [0]{\spacefactor3000\relax}%
\providecommand \BibitemShut  [1]{\csname bibitem#1\endcsname}%
\let\auto@bib@innerbib\@empty
%</preamble>
\bibitem [{\citenamefont {Moore}\ \emph {et~al.}(2019)\citenamefont {Moore},
  \citenamefont {Boland}, \citenamefont {Camara}, \citenamefont {Chervitz},
  \citenamefont {Gonzalez}, \citenamefont {Himes}, \citenamefont {Kim},
  \citenamefont {Mowery}, \citenamefont {Ritchie}, \citenamefont {Shen},
  \citenamefont {Urbanowicz},\ and\ \citenamefont
  {Holmes}}]{moore2019preparing}%
  \BibitemOpen
  \bibfield  {author} {\bibinfo {author} {\bibfnamefont {J.~H.}\ \bibnamefont
  {Moore}}, \bibinfo {author} {\bibfnamefont {M.~R.}\ \bibnamefont {Boland}},
  \bibinfo {author} {\bibfnamefont {P.~G.}\ \bibnamefont {Camara}}, \bibinfo
  {author} {\bibfnamefont {H.}~\bibnamefont {Chervitz}}, \bibinfo {author}
  {\bibfnamefont {G.}~\bibnamefont {Gonzalez}}, \bibinfo {author}
  {\bibfnamefont {B.~E.}\ \bibnamefont {Himes}}, \bibinfo {author}
  {\bibfnamefont {D.}~\bibnamefont {Kim}}, \bibinfo {author} {\bibfnamefont
  {D.~L.}\ \bibnamefont {Mowery}}, \bibinfo {author} {\bibfnamefont {M.~D.}\
  \bibnamefont {Ritchie}}, \bibinfo {author} {\bibfnamefont {L.}~\bibnamefont
  {Shen}}, \bibinfo {author} {\bibfnamefont {R.~J.}\ \bibnamefont
  {Urbanowicz}},\ and\ \bibinfo {author} {\bibfnamefont {J.~H.}\ \bibnamefont
  {Holmes}},\ }\bibfield  {title} {\bibinfo {title} {Preparing next-generation
  scientists for biomedical big data: artificial intelligence approaches},\
  }\href {https://doi.org/10.2217/pme-2018-0145} {\bibfield  {journal}
  {\bibinfo  {journal} {Personalized Medicine}\ }\textbf {\bibinfo {volume}
  {16}},\ \bibinfo {pages} {247} (\bibinfo {year} {2019})}\BibitemShut
  {NoStop}%
\bibitem [{\citenamefont {He}\ \emph {et~al.}(2019)\citenamefont {He},
  \citenamefont {Baxter}, \citenamefont {Xu}, \citenamefont {Xu}, \citenamefont
  {Zhou},\ and\ \citenamefont {Zhang}}]{he2019practical}%
  \BibitemOpen
  \bibfield  {author} {\bibinfo {author} {\bibfnamefont {J.}~\bibnamefont
  {He}}, \bibinfo {author} {\bibfnamefont {S.~L.}\ \bibnamefont {Baxter}},
  \bibinfo {author} {\bibfnamefont {J.}~\bibnamefont {Xu}}, \bibinfo {author}
  {\bibfnamefont {J.}~\bibnamefont {Xu}}, \bibinfo {author} {\bibfnamefont
  {X.}~\bibnamefont {Zhou}},\ and\ \bibinfo {author} {\bibfnamefont
  {K.}~\bibnamefont {Zhang}},\ }\bibfield  {title} {\bibinfo {title} {The
  practical implementation of artificial intelligence technologies in
  medicine},\ }\href {https://doi.org/10.1038/s41591-018-0307-0} {\bibfield
  {journal} {\bibinfo  {journal} {Nature Medicine}\ }\textbf {\bibinfo {volume}
  {25}},\ \bibinfo {pages} {30} (\bibinfo {year} {2019})}\BibitemShut {NoStop}%
\bibitem [{\citenamefont {Hudson}\ and\ \citenamefont
  {Cohen}(2000)}]{hudson2000neural}%
  \BibitemOpen
  \bibfield  {author} {\bibinfo {author} {\bibfnamefont {D.~L.}\ \bibnamefont
  {Hudson}}\ and\ \bibinfo {author} {\bibfnamefont {M.~E.}\ \bibnamefont
  {Cohen}},\ }\href {https://doi.org/10.1109/9780470545355} {\emph {\bibinfo
  {title} {Neural networks and artificial intelligence for biomedical
  engineering}}}\ (\bibinfo  {publisher} {Wiley Online Library},\ \bibinfo
  {year} {2000})\BibitemShut {NoStop}%
\bibitem [{\citenamefont {Febbraro}\ \emph {et~al.}(2022)\citenamefont
  {Febbraro}, \citenamefont {Gheware}, \citenamefont {Kennedy}, \citenamefont
  {Jain}, \citenamefont {de~Moraes},\ and\ \citenamefont
  {Juergens}}]{febbraro2022barriers}%
  \BibitemOpen
  \bibfield  {author} {\bibinfo {author} {\bibfnamefont {M.}~\bibnamefont
  {Febbraro}}, \bibinfo {author} {\bibfnamefont {A.}~\bibnamefont {Gheware}},
  \bibinfo {author} {\bibfnamefont {T.}~\bibnamefont {Kennedy}}, \bibinfo
  {author} {\bibfnamefont {D.}~\bibnamefont {Jain}}, \bibinfo {author}
  {\bibfnamefont {F.~Y.}\ \bibnamefont {de~Moraes}},\ and\ \bibinfo {author}
  {\bibfnamefont {R.}~\bibnamefont {Juergens}},\ }\bibfield  {title} {\bibinfo
  {title} {Barriers to access: Global variability in implementing treatment
  advances in lung cancer},\ }\href {https://doi.org/10.1200/EDBK\_351021}
  {\bibfield  {journal} {\bibinfo  {journal} {American Society of Clinical
  Oncology Educational Book}\ ,\ \bibinfo {pages} {666}} (\bibinfo {year}
  {2022})},\ \bibinfo {note} {pMID: 35427189}\BibitemShut {NoStop}%
\bibitem [{\citenamefont {Jumper}\ \emph {et~al.}(2021)\citenamefont {Jumper},
  \citenamefont {Evans}, \citenamefont {Pritzel}, \citenamefont {Green},
  \citenamefont {Figurnov}, \citenamefont {Ronneberger}, \citenamefont
  {Tunyasuvunakool}, \citenamefont {Bates}, \citenamefont {{\v{Z}}{\'\i}dek},
  \citenamefont {Potapenko} \emph {et~al.}}]{jumper2021highly}%
  \BibitemOpen
  \bibfield  {author} {\bibinfo {author} {\bibfnamefont {J.}~\bibnamefont
  {Jumper}}, \bibinfo {author} {\bibfnamefont {R.}~\bibnamefont {Evans}},
  \bibinfo {author} {\bibfnamefont {A.}~\bibnamefont {Pritzel}}, \bibinfo
  {author} {\bibfnamefont {T.}~\bibnamefont {Green}}, \bibinfo {author}
  {\bibfnamefont {M.}~\bibnamefont {Figurnov}}, \bibinfo {author}
  {\bibfnamefont {O.}~\bibnamefont {Ronneberger}}, \bibinfo {author}
  {\bibfnamefont {K.}~\bibnamefont {Tunyasuvunakool}}, \bibinfo {author}
  {\bibfnamefont {R.}~\bibnamefont {Bates}}, \bibinfo {author} {\bibfnamefont
  {A.}~\bibnamefont {{\v{Z}}{\'\i}dek}}, \bibinfo {author} {\bibfnamefont
  {A.}~\bibnamefont {Potapenko}}, \emph {et~al.},\ }\bibfield  {title}
  {\bibinfo {title} {Highly accurate protein structure prediction with
  alphafold},\ }\href {https://doi.org/10.1038/s41586-021-03819-2} {\bibfield
  {journal} {\bibinfo  {journal} {Nature}\ }\textbf {\bibinfo {volume} {596}},\
  \bibinfo {pages} {583} (\bibinfo {year} {2021})}\BibitemShut {NoStop}%
\bibitem [{\citenamefont {Tunyasuvunakool}\ \emph {et~al.}(2021)\citenamefont
  {Tunyasuvunakool}, \citenamefont {Adler}, \citenamefont {Wu}, \citenamefont
  {Green}, \citenamefont {Zielinski}, \citenamefont {{\v{Z}}{\'\i}dek},
  \citenamefont {Bridgland}, \citenamefont {Cowie}, \citenamefont {Meyer},
  \citenamefont {Laydon} \emph {et~al.}}]{tunyasuvunakool2021highly}%
  \BibitemOpen
  \bibfield  {author} {\bibinfo {author} {\bibfnamefont {K.}~\bibnamefont
  {Tunyasuvunakool}}, \bibinfo {author} {\bibfnamefont {J.}~\bibnamefont
  {Adler}}, \bibinfo {author} {\bibfnamefont {Z.}~\bibnamefont {Wu}}, \bibinfo
  {author} {\bibfnamefont {T.}~\bibnamefont {Green}}, \bibinfo {author}
  {\bibfnamefont {M.}~\bibnamefont {Zielinski}}, \bibinfo {author}
  {\bibfnamefont {A.}~\bibnamefont {{\v{Z}}{\'\i}dek}}, \bibinfo {author}
  {\bibfnamefont {A.}~\bibnamefont {Bridgland}}, \bibinfo {author}
  {\bibfnamefont {A.}~\bibnamefont {Cowie}}, \bibinfo {author} {\bibfnamefont
  {C.}~\bibnamefont {Meyer}}, \bibinfo {author} {\bibfnamefont
  {A.}~\bibnamefont {Laydon}}, \emph {et~al.},\ }\bibfield  {title} {\bibinfo
  {title} {Highly accurate protein structure prediction for the human
  proteome},\ }\href {https://doi.org/10.1038/s41586-021-03828-1} {\bibfield
  {journal} {\bibinfo  {journal} {Nature}\ }\textbf {\bibinfo {volume} {596}},\
  \bibinfo {pages} {590} (\bibinfo {year} {2021})}\BibitemShut {NoStop}%
\bibitem [{\citenamefont {Zoabi}\ \emph {et~al.}(2021)\citenamefont {Zoabi},
  \citenamefont {Deri-Rozov},\ and\ \citenamefont
  {Shomron}}]{zoabi2021machine}%
  \BibitemOpen
  \bibfield  {author} {\bibinfo {author} {\bibfnamefont {Y.}~\bibnamefont
  {Zoabi}}, \bibinfo {author} {\bibfnamefont {S.}~\bibnamefont {Deri-Rozov}},\
  and\ \bibinfo {author} {\bibfnamefont {N.}~\bibnamefont {Shomron}},\
  }\bibfield  {title} {\bibinfo {title} {Machine learning-based prediction of
  covid-19 diagnosis based on symptoms},\ }\href
  {https://doi.org/10.1038/s41746-020-00372-6} {\bibfield  {journal} {\bibinfo
  {journal} {npj digital medicine}\ }\textbf {\bibinfo {volume} {4}},\ \bibinfo
  {pages} {1} (\bibinfo {year} {2021})}\BibitemShut {NoStop}%
\bibitem [{\citenamefont {Réda}\ \emph {et~al.}(2020)\citenamefont {Réda},
  \citenamefont {Kaufmann},\ and\ \citenamefont
  {Delahaye-Duriez}}]{reda2020machine}%
  \BibitemOpen
  \bibfield  {author} {\bibinfo {author} {\bibfnamefont {C.}~\bibnamefont
  {Réda}}, \bibinfo {author} {\bibfnamefont {E.}~\bibnamefont {Kaufmann}},\
  and\ \bibinfo {author} {\bibfnamefont {A.}~\bibnamefont {Delahaye-Duriez}},\
  }\bibfield  {title} {\bibinfo {title} {Machine learning applications in drug
  development},\ }\href
  {https://doi.org/https://doi.org/10.1016/j.csbj.2019.12.006} {\bibfield
  {journal} {\bibinfo  {journal} {Computational and Structural Biotechnology
  Journal}\ }\textbf {\bibinfo {volume} {18}},\ \bibinfo {pages} {241}
  (\bibinfo {year} {2020})}\BibitemShut {NoStop}%
\bibitem [{\citenamefont {Ekins}\ \emph {et~al.}(2019)\citenamefont {Ekins},
  \citenamefont {Puhl}, \citenamefont {Zorn}, \citenamefont {Lane},
  \citenamefont {Russo}, \citenamefont {Klein}, \citenamefont {Hickey},\ and\
  \citenamefont {Clark}}]{ekins2019exploiting}%
  \BibitemOpen
  \bibfield  {author} {\bibinfo {author} {\bibfnamefont {S.}~\bibnamefont
  {Ekins}}, \bibinfo {author} {\bibfnamefont {A.~C.}\ \bibnamefont {Puhl}},
  \bibinfo {author} {\bibfnamefont {K.~M.}\ \bibnamefont {Zorn}}, \bibinfo
  {author} {\bibfnamefont {T.~R.}\ \bibnamefont {Lane}}, \bibinfo {author}
  {\bibfnamefont {D.~P.}\ \bibnamefont {Russo}}, \bibinfo {author}
  {\bibfnamefont {J.~J.}\ \bibnamefont {Klein}}, \bibinfo {author}
  {\bibfnamefont {A.~J.}\ \bibnamefont {Hickey}},\ and\ \bibinfo {author}
  {\bibfnamefont {A.~M.}\ \bibnamefont {Clark}},\ }\bibfield  {title} {\bibinfo
  {title} {Exploiting machine learning for end-to-end drug discovery and
  development},\ }\href {https://doi.org/10.1038/s41563-019-0338-z} {\bibfield
  {journal} {\bibinfo  {journal} {Nature materials}\ }\textbf {\bibinfo
  {volume} {18}},\ \bibinfo {pages} {435} (\bibinfo {year} {2019})}\BibitemShut
  {NoStop}%
\bibitem [{\citenamefont {Shen}\ \emph {et~al.}(2017)\citenamefont {Shen},
  \citenamefont {Wu},\ and\ \citenamefont {Suk}}]{shen2017deep}%
  \BibitemOpen
  \bibfield  {author} {\bibinfo {author} {\bibfnamefont {D.}~\bibnamefont
  {Shen}}, \bibinfo {author} {\bibfnamefont {G.}~\bibnamefont {Wu}},\ and\
  \bibinfo {author} {\bibfnamefont {H.-I.}\ \bibnamefont {Suk}},\ }\bibfield
  {title} {\bibinfo {title} {Deep learning in medical image analysis},\ }\href
  {https://doi.org/10.1146/annurev-bioeng-071516-044442} {\bibfield  {journal}
  {\bibinfo  {journal} {Annual review of biomedical engineering}\ }\textbf
  {\bibinfo {volume} {19}},\ \bibinfo {pages} {221} (\bibinfo {year}
  {2017})}\BibitemShut {NoStop}%
\bibitem [{\citenamefont {Liu}\ \emph {et~al.}(2017)\citenamefont {Liu},
  \citenamefont {Wang}, \citenamefont {Du},\ and\ \citenamefont
  {Jing}}]{liu2017raman}%
  \BibitemOpen
  \bibfield  {author} {\bibinfo {author} {\bibfnamefont {W.}~\bibnamefont
  {Liu}}, \bibinfo {author} {\bibfnamefont {H.}~\bibnamefont {Wang}}, \bibinfo
  {author} {\bibfnamefont {J.}~\bibnamefont {Du}},\ and\ \bibinfo {author}
  {\bibfnamefont {C.}~\bibnamefont {Jing}},\ }\bibfield  {title} {\bibinfo
  {title} {Raman microspectroscopy of nucleus and cytoplasm for human colon
  cancer diagnosis},\ }\href
  {https://doi.org/https://doi.org/10.1016/j.bios.2017.05.045} {\bibfield
  {journal} {\bibinfo  {journal} {Biosensors and Bioelectronics}\ }\textbf
  {\bibinfo {volume} {97}},\ \bibinfo {pages} {70} (\bibinfo {year}
  {2017})}\BibitemShut {NoStop}%
\bibitem [{\citenamefont {Bahreini}\ \emph {et~al.}(2019)\citenamefont
  {Bahreini}, \citenamefont {Hosseinzadegan}, \citenamefont {Rashidi},
  \citenamefont {Miri}, \citenamefont {Mirzaei},\ and\ \citenamefont
  {Hajian}}]{bahreini2019raman}%
  \BibitemOpen
  \bibfield  {author} {\bibinfo {author} {\bibfnamefont {M.}~\bibnamefont
  {Bahreini}}, \bibinfo {author} {\bibfnamefont {A.}~\bibnamefont
  {Hosseinzadegan}}, \bibinfo {author} {\bibfnamefont {A.}~\bibnamefont
  {Rashidi}}, \bibinfo {author} {\bibfnamefont {S.~R.}\ \bibnamefont {Miri}},
  \bibinfo {author} {\bibfnamefont {H.~R.}\ \bibnamefont {Mirzaei}},\ and\
  \bibinfo {author} {\bibfnamefont {P.}~\bibnamefont {Hajian}},\ }\bibfield
  {title} {\bibinfo {title} {A raman-based serum constituents’ analysis for
  gastric cancer diagnosis: In vitro study},\ }\href
  {https://doi.org/https://doi.org/10.1016/j.talanta.2019.06.068} {\bibfield
  {journal} {\bibinfo  {journal} {Talanta}\ }\textbf {\bibinfo {volume}
  {204}},\ \bibinfo {pages} {826} (\bibinfo {year} {2019})}\BibitemShut
  {NoStop}%
\bibitem [{\citenamefont {Z{\'u}{\~{n}}iga}\ \emph {et~al.}(2019)\citenamefont
  {Z{\'u}{\~{n}}iga}, \citenamefont {Jones}, \citenamefont {Anderson},
  \citenamefont {Echevarria}, \citenamefont {Miller}, \citenamefont {Stashko},
  \citenamefont {Schmolze}, \citenamefont {Cha}, \citenamefont {Kothari},
  \citenamefont {Fong},\ and\ \citenamefont
  {Storrie-Lombardi}}]{zuniga2019raman}%
  \BibitemOpen
  \bibfield  {author} {\bibinfo {author} {\bibfnamefont {W.~C.}\ \bibnamefont
  {Z{\'u}{\~{n}}iga}}, \bibinfo {author} {\bibfnamefont {V.}~\bibnamefont
  {Jones}}, \bibinfo {author} {\bibfnamefont {S.~M.}\ \bibnamefont {Anderson}},
  \bibinfo {author} {\bibfnamefont {A.}~\bibnamefont {Echevarria}}, \bibinfo
  {author} {\bibfnamefont {N.~L.}\ \bibnamefont {Miller}}, \bibinfo {author}
  {\bibfnamefont {C.}~\bibnamefont {Stashko}}, \bibinfo {author} {\bibfnamefont
  {D.}~\bibnamefont {Schmolze}}, \bibinfo {author} {\bibfnamefont {P.~D.}\
  \bibnamefont {Cha}}, \bibinfo {author} {\bibfnamefont {R.}~\bibnamefont
  {Kothari}}, \bibinfo {author} {\bibfnamefont {Y.}~\bibnamefont {Fong}},\ and\
  \bibinfo {author} {\bibfnamefont {M.~C.}\ \bibnamefont {Storrie-Lombardi}},\
  }\bibfield  {title} {\bibinfo {title} {Raman spectroscopy for rapid
  evaluation of surgical margins during breast cancer lumpectomy},\ }\href
  {https://doi.org/10.1038/s41598-019-51112-0} {\bibfield  {journal} {\bibinfo
  {journal} {Scientific Reports}\ }\textbf {\bibinfo {volume} {9}},\ \bibinfo
  {pages} {14639} (\bibinfo {year} {2019})}\BibitemShut {NoStop}%
\bibitem [{\citenamefont {Gilpin}\ \emph {et~al.}(2018)\citenamefont {Gilpin},
  \citenamefont {Bau}, \citenamefont {Yuan}, \citenamefont {Bajwa},
  \citenamefont {Specter},\ and\ \citenamefont {Kagal}}]{gilpin2018explaining}%
  \BibitemOpen
  \bibfield  {author} {\bibinfo {author} {\bibfnamefont {L.~H.}\ \bibnamefont
  {Gilpin}}, \bibinfo {author} {\bibfnamefont {D.}~\bibnamefont {Bau}},
  \bibinfo {author} {\bibfnamefont {B.~Z.}\ \bibnamefont {Yuan}}, \bibinfo
  {author} {\bibfnamefont {A.}~\bibnamefont {Bajwa}}, \bibinfo {author}
  {\bibfnamefont {M.}~\bibnamefont {Specter}},\ and\ \bibinfo {author}
  {\bibfnamefont {L.}~\bibnamefont {Kagal}},\ }\bibfield  {title} {\bibinfo
  {title} {Explaining explanations: An overview of interpretability of machine
  learning},\ }in\ \href {https://doi.org/10.1109/DSAA.2018.00018} {\emph
  {\bibinfo {booktitle} {2018 IEEE 5th International Conference on data science
  and advanced analytics (DSAA)}}}\ (\bibinfo {organization} {IEEE},\ \bibinfo
  {year} {2018})\ pp.\ \bibinfo {pages} {80--89}\BibitemShut {NoStop}%
\bibitem [{\citenamefont {Zhang}\ and\ \citenamefont
  {Zhu}(2018)}]{zhang2018visual}%
  \BibitemOpen
  \bibfield  {author} {\bibinfo {author} {\bibfnamefont {Q.-s.}\ \bibnamefont
  {Zhang}}\ and\ \bibinfo {author} {\bibfnamefont {S.-C.}\ \bibnamefont
  {Zhu}},\ }\bibfield  {title} {\bibinfo {title} {Visual interpretability for
  deep learning: a survey},\ }\href {https://doi.org/10.1631/FITEE.1700808}
  {\bibfield  {journal} {\bibinfo  {journal} {Frontiers of Information
  Technology \& Electronic Engineering}\ }\textbf {\bibinfo {volume} {19}},\
  \bibinfo {pages} {27} (\bibinfo {year} {2018})}\BibitemShut {NoStop}%
\bibitem [{\citenamefont {Carvalho}\ \emph {et~al.}(2019)\citenamefont
  {Carvalho}, \citenamefont {Pereira},\ and\ \citenamefont
  {Cardoso}}]{carvalho2019machine}%
  \BibitemOpen
  \bibfield  {author} {\bibinfo {author} {\bibfnamefont {D.~V.}\ \bibnamefont
  {Carvalho}}, \bibinfo {author} {\bibfnamefont {E.~M.}\ \bibnamefont
  {Pereira}},\ and\ \bibinfo {author} {\bibfnamefont {J.~S.}\ \bibnamefont
  {Cardoso}},\ }\bibfield  {title} {\bibinfo {title} {Machine learning
  interpretability: A survey on methods and metrics},\ }\bibfield  {journal}
  {\bibinfo  {journal} {Electronics}\ }\textbf {\bibinfo {volume} {8}},\ \href
  {https://doi.org/10.3390/electronics8080832} {10.3390/electronics8080832}
  (\bibinfo {year} {2019})\BibitemShut {NoStop}%
\bibitem [{\citenamefont {Stoudenmire}\ and\ \citenamefont
  {Schwab}(2016)}]{SS16TNML}%
  \BibitemOpen
  \bibfield  {author} {\bibinfo {author} {\bibfnamefont {E.}~\bibnamefont
  {Stoudenmire}}\ and\ \bibinfo {author} {\bibfnamefont {D.~J.}\ \bibnamefont
  {Schwab}},\ }\bibfield  {title} {\bibinfo {title} {Supervised learning with
  tensor networks},\ }in\ \href
  {https://proceedings.neurips.cc/paper/2016/file/5314b9674c86e3f9d1ba25ef9bb32895-Paper.pdf}
  {\emph {\bibinfo {booktitle} {Advances in Neural Information Processing
  Systems}}},\ Vol.~\bibinfo {volume} {29},\ \bibinfo {editor} {edited by\
  \bibinfo {editor} {\bibfnamefont {D.}~\bibnamefont {Lee}}, \bibinfo {editor}
  {\bibfnamefont {M.}~\bibnamefont {Sugiyama}}, \bibinfo {editor}
  {\bibfnamefont {U.}~\bibnamefont {Luxburg}}, \bibinfo {editor} {\bibfnamefont
  {I.}~\bibnamefont {Guyon}},\ and\ \bibinfo {editor} {\bibfnamefont
  {R.}~\bibnamefont {Garnett}}}\ (\bibinfo  {publisher} {Curran Associates,
  Inc.},\ \bibinfo {year} {2016})\BibitemShut {NoStop}%
\bibitem [{\citenamefont {Liu}\ \emph {et~al.}(2019)\citenamefont {Liu},
  \citenamefont {Ran}, \citenamefont {Wittek}, \citenamefont {Peng},
  \citenamefont {Garc{\'{\i}}a}, \citenamefont {Su},\ and\ \citenamefont
  {Lewenstein}}]{liu2019machine}%
  \BibitemOpen
  \bibfield  {author} {\bibinfo {author} {\bibfnamefont {D.}~\bibnamefont
  {Liu}}, \bibinfo {author} {\bibfnamefont {S.-J.}\ \bibnamefont {Ran}},
  \bibinfo {author} {\bibfnamefont {P.}~\bibnamefont {Wittek}}, \bibinfo
  {author} {\bibfnamefont {C.}~\bibnamefont {Peng}}, \bibinfo {author}
  {\bibfnamefont {R.~B.}\ \bibnamefont {Garc{\'{\i}}a}}, \bibinfo {author}
  {\bibfnamefont {G.}~\bibnamefont {Su}},\ and\ \bibinfo {author}
  {\bibfnamefont {M.}~\bibnamefont {Lewenstein}},\ }\bibfield  {title}
  {\bibinfo {title} {Machine learning by unitary tensor network of hierarchical
  tree structure},\ }\href {https://doi.org/10.1088/1367-2630/ab31ef}
  {\bibfield  {journal} {\bibinfo  {journal} {New Journal of Physics}\ }\textbf
  {\bibinfo {volume} {21}},\ \bibinfo {pages} {073059} (\bibinfo {year}
  {2019})}\BibitemShut {NoStop}%
\bibitem [{\citenamefont {Han}\ \emph {et~al.}(2018)\citenamefont {Han},
  \citenamefont {Wang}, \citenamefont {Fan}, \citenamefont {Wang},\ and\
  \citenamefont {Zhang}}]{han2018unsupervised}%
  \BibitemOpen
  \bibfield  {author} {\bibinfo {author} {\bibfnamefont {Z.-Y.}\ \bibnamefont
  {Han}}, \bibinfo {author} {\bibfnamefont {J.}~\bibnamefont {Wang}}, \bibinfo
  {author} {\bibfnamefont {H.}~\bibnamefont {Fan}}, \bibinfo {author}
  {\bibfnamefont {L.}~\bibnamefont {Wang}},\ and\ \bibinfo {author}
  {\bibfnamefont {P.}~\bibnamefont {Zhang}},\ }\bibfield  {title} {\bibinfo
  {title} {Unsupervised generative modeling using matrix product states},\
  }\href {https://doi.org/10.1103/PhysRevX.8.031012} {\bibfield  {journal}
  {\bibinfo  {journal} {Phys. Rev. X}\ }\textbf {\bibinfo {volume} {8}},\
  \bibinfo {pages} {031012} (\bibinfo {year} {2018})}\BibitemShut {NoStop}%
\bibitem [{\citenamefont {Cheng}\ \emph {et~al.}(2019)\citenamefont {Cheng},
  \citenamefont {Wang}, \citenamefont {Xiang},\ and\ \citenamefont
  {Zhang}}]{cheng2019tree}%
  \BibitemOpen
  \bibfield  {author} {\bibinfo {author} {\bibfnamefont {S.}~\bibnamefont
  {Cheng}}, \bibinfo {author} {\bibfnamefont {L.}~\bibnamefont {Wang}},
  \bibinfo {author} {\bibfnamefont {T.}~\bibnamefont {Xiang}},\ and\ \bibinfo
  {author} {\bibfnamefont {P.}~\bibnamefont {Zhang}},\ }\bibfield  {title}
  {\bibinfo {title} {Tree tensor networks for generative modeling},\ }\href
  {https://doi.org/10.1103/PhysRevB.99.155131} {\bibfield  {journal} {\bibinfo
  {journal} {Phys. Rev. B}\ }\textbf {\bibinfo {volume} {99}},\ \bibinfo
  {pages} {155131} (\bibinfo {year} {2019})}\BibitemShut {NoStop}%
\bibitem [{\citenamefont {Vieijra}\ \emph {et~al.}(2022)\citenamefont
  {Vieijra}, \citenamefont {Vanderstraeten},\ and\ \citenamefont
  {Verstraete}}]{vieijra2022generative}%
  \BibitemOpen
  \bibfield  {author} {\bibinfo {author} {\bibfnamefont {T.}~\bibnamefont
  {Vieijra}}, \bibinfo {author} {\bibfnamefont {L.}~\bibnamefont
  {Vanderstraeten}},\ and\ \bibinfo {author} {\bibfnamefont {F.}~\bibnamefont
  {Verstraete}},\ }\bibfield  {title} {\bibinfo {title} {Generative modeling
  with projected entangled-pair states},\ }\href@noop {} {\bibfield  {journal}
  {\bibinfo  {journal} {arXiv preprint arXiv:2202.08177}\ } (\bibinfo {year}
  {2022})}\BibitemShut {NoStop}%
\bibitem [{\citenamefont {Sun}\ \emph {et~al.}(2020)\citenamefont {Sun},
  \citenamefont {Peng}, \citenamefont {Liu}, \citenamefont {Ran},\ and\
  \citenamefont {Su}}]{sun2020generative}%
  \BibitemOpen
  \bibfield  {author} {\bibinfo {author} {\bibfnamefont {Z.-Z.}\ \bibnamefont
  {Sun}}, \bibinfo {author} {\bibfnamefont {C.}~\bibnamefont {Peng}}, \bibinfo
  {author} {\bibfnamefont {D.}~\bibnamefont {Liu}}, \bibinfo {author}
  {\bibfnamefont {S.-J.}\ \bibnamefont {Ran}},\ and\ \bibinfo {author}
  {\bibfnamefont {G.}~\bibnamefont {Su}},\ }\bibfield  {title} {\bibinfo
  {title} {Generative tensor network classification model for supervised
  machine learning},\ }\href {https://doi.org/10.1103/PhysRevB.101.075135}
  {\bibfield  {journal} {\bibinfo  {journal} {Phys. Rev. B}\ }\textbf {\bibinfo
  {volume} {101}},\ \bibinfo {pages} {075135} (\bibinfo {year}
  {2020})}\BibitemShut {NoStop}%
\bibitem [{\citenamefont {Wang}\ \emph {et~al.}(2020)\citenamefont {Wang},
  \citenamefont {Roberts}, \citenamefont {Vidal},\ and\ \citenamefont
  {Leichenauer}}]{wang2020anomaly}%
  \BibitemOpen
  \bibfield  {author} {\bibinfo {author} {\bibfnamefont {J.}~\bibnamefont
  {Wang}}, \bibinfo {author} {\bibfnamefont {C.}~\bibnamefont {Roberts}},
  \bibinfo {author} {\bibfnamefont {G.}~\bibnamefont {Vidal}},\ and\ \bibinfo
  {author} {\bibfnamefont {S.}~\bibnamefont {Leichenauer}},\ }\href@noop {}
  {\bibinfo {title} {Anomaly detection with tensor networks}} (\bibinfo {year}
  {2020}),\ \Eprint {https://arxiv.org/abs/2006.02516} {arXiv:2006.02516
  [cs.LG]} \BibitemShut {NoStop}%
\bibitem [{\citenamefont {Ran}\ \emph {et~al.}(2020{\natexlab{a}})\citenamefont
  {Ran}, \citenamefont {Tirrito}, \citenamefont {Peng}, \citenamefont {Chen},
  \citenamefont {Tagliacozzo}, \citenamefont {Su},\ and\ \citenamefont
  {Lewenstein}}]{ran2020tensor}%
  \BibitemOpen
  \bibfield  {author} {\bibinfo {author} {\bibfnamefont {S.-J.}\ \bibnamefont
  {Ran}}, \bibinfo {author} {\bibfnamefont {E.}~\bibnamefont {Tirrito}},
  \bibinfo {author} {\bibfnamefont {C.}~\bibnamefont {Peng}}, \bibinfo {author}
  {\bibfnamefont {X.}~\bibnamefont {Chen}}, \bibinfo {author} {\bibfnamefont
  {L.}~\bibnamefont {Tagliacozzo}}, \bibinfo {author} {\bibfnamefont
  {G.}~\bibnamefont {Su}},\ and\ \bibinfo {author} {\bibfnamefont
  {M.}~\bibnamefont {Lewenstein}},\ }\href
  {https://doi.org/10.1007/978-3-030-34489-4} {\emph {\bibinfo {title} {Tensor
  Network Contractions: Methods and Applications to Quantum Many-Body
  Systems}}}\ (\bibinfo  {publisher} {Springer, Cham},\ \bibinfo {year}
  {2020})\BibitemShut {NoStop}%
\bibitem [{\citenamefont {Verstraete}\ \emph {et~al.}(2008)\citenamefont
  {Verstraete}, \citenamefont {Murg},\ and\ \citenamefont
  {Cirac}}]{verstraete2008matrix}%
  \BibitemOpen
  \bibfield  {author} {\bibinfo {author} {\bibfnamefont {F.}~\bibnamefont
  {Verstraete}}, \bibinfo {author} {\bibfnamefont {V.}~\bibnamefont {Murg}},\
  and\ \bibinfo {author} {\bibfnamefont {J.}~\bibnamefont {Cirac}},\ }\bibfield
   {title} {\bibinfo {title} {Matrix product states, projected entangled pair
  states, and variational renormalization group methods for quantum spin
  systems},\ }\href {https://doi.org/10.1080/14789940801912366} {\bibfield
  {journal} {\bibinfo  {journal} {Advances in Physics}\ }\textbf {\bibinfo
  {volume} {57}},\ \bibinfo {pages} {143} (\bibinfo {year} {2008})}\BibitemShut
  {NoStop}%
\bibitem [{\citenamefont {Or{\'u}s}(2019)}]{orus2019tensor}%
  \BibitemOpen
  \bibfield  {author} {\bibinfo {author} {\bibfnamefont {R.}~\bibnamefont
  {Or{\'u}s}},\ }\bibfield  {title} {\bibinfo {title} {Tensor networks for
  complex quantum systems},\ }\href {https://doi.org/10.1038/s42254-019-0086-7}
  {\bibfield  {journal} {\bibinfo  {journal} {Nature Reviews Physics}\ }\textbf
  {\bibinfo {volume} {1}},\ \bibinfo {pages} {538} (\bibinfo {year}
  {2019})}\BibitemShut {NoStop}%
\bibitem [{\citenamefont {Cirac}\ \emph {et~al.}(2021)\citenamefont {Cirac},
  \citenamefont {P\'erez-Garc\'{\i}a}, \citenamefont {Schuch},\ and\
  \citenamefont {Verstraete}}]{cirac2021matrix}%
  \BibitemOpen
  \bibfield  {author} {\bibinfo {author} {\bibfnamefont {J.~I.}\ \bibnamefont
  {Cirac}}, \bibinfo {author} {\bibfnamefont {D.}~\bibnamefont
  {P\'erez-Garc\'{\i}a}}, \bibinfo {author} {\bibfnamefont {N.}~\bibnamefont
  {Schuch}},\ and\ \bibinfo {author} {\bibfnamefont {F.}~\bibnamefont
  {Verstraete}},\ }\bibfield  {title} {\bibinfo {title} {Matrix product states
  and projected entangled pair states: Concepts, symmetries, theorems},\ }\href
  {https://doi.org/10.1103/RevModPhys.93.045003} {\bibfield  {journal}
  {\bibinfo  {journal} {Rev. Mod. Phys.}\ }\textbf {\bibinfo {volume} {93}},\
  \bibinfo {pages} {045003} (\bibinfo {year} {2021})}\BibitemShut {NoStop}%
\bibitem [{\citenamefont {Liu}\ \emph {et~al.}(2021)\citenamefont {Liu},
  \citenamefont {Li}, \citenamefont {Zhang}, \citenamefont {Lewenstein},
  \citenamefont {Su},\ and\ \citenamefont {Ran}}]{liu2021entanglement}%
  \BibitemOpen
  \bibfield  {author} {\bibinfo {author} {\bibfnamefont {Y.}~\bibnamefont
  {Liu}}, \bibinfo {author} {\bibfnamefont {W.-J.}\ \bibnamefont {Li}},
  \bibinfo {author} {\bibfnamefont {X.}~\bibnamefont {Zhang}}, \bibinfo
  {author} {\bibfnamefont {M.}~\bibnamefont {Lewenstein}}, \bibinfo {author}
  {\bibfnamefont {G.}~\bibnamefont {Su}},\ and\ \bibinfo {author}
  {\bibfnamefont {S.-J.}\ \bibnamefont {Ran}},\ }\bibfield  {title} {\bibinfo
  {title} {Entanglement-based feature extraction by tensor network machine
  learning},\ }\bibfield  {journal} {\bibinfo  {journal} {Frontiers in Applied
  Mathematics and Statistics}\ }\textbf {\bibinfo {volume} {7}},\ \href
  {https://doi.org/10.3389/fams.2021.716044} {10.3389/fams.2021.716044}
  (\bibinfo {year} {2021})\BibitemShut {NoStop}%
\bibitem [{\citenamefont {Bai}\ \emph {et~al.}(2022)\citenamefont {Bai},
  \citenamefont {Tang},\ and\ \citenamefont {Ran}}]{Bai:100701}%
  \BibitemOpen
  \bibfield  {author} {\bibinfo {author} {\bibfnamefont {S.-C.}\ \bibnamefont
  {Bai}}, \bibinfo {author} {\bibfnamefont {Y.-C.}\ \bibnamefont {Tang}},\ and\
  \bibinfo {author} {\bibfnamefont {S.-J.}\ \bibnamefont {Ran}},\ }\bibfield
  {title} {\bibinfo {title} {Unsupervised recognition of informative features
  via tensor network machine learning and quantum entanglement variations},\
  }\href {https://doi.org/10.1088/0256-307X/39/10/100701} {\bibfield  {journal}
  {\bibinfo  {journal} {Chinese Physics Letters}\ }\textbf {\bibinfo {volume}
  {39}},\ \bibinfo {eid} {100701} (\bibinfo {year} {2022})}\BibitemShut
  {NoStop}%
\bibitem [{\citenamefont {Ran}\ \emph {et~al.}(2020{\natexlab{b}})\citenamefont
  {Ran}, \citenamefont {Sun}, \citenamefont {Fei}, \citenamefont {Su},\ and\
  \citenamefont {Lewenstein}}]{RSF+20TNCS}%
  \BibitemOpen
  \bibfield  {author} {\bibinfo {author} {\bibfnamefont {S.-J.}\ \bibnamefont
  {Ran}}, \bibinfo {author} {\bibfnamefont {Z.-Z.}\ \bibnamefont {Sun}},
  \bibinfo {author} {\bibfnamefont {S.-M.}\ \bibnamefont {Fei}}, \bibinfo
  {author} {\bibfnamefont {G.}~\bibnamefont {Su}},\ and\ \bibinfo {author}
  {\bibfnamefont {M.}~\bibnamefont {Lewenstein}},\ }\bibfield  {title}
  {\bibinfo {title} {Tensor network compressed sensing with unsupervised
  machine learning},\ }\href {https://doi.org/10.1103/PhysRevResearch.2.033293}
  {\bibfield  {journal} {\bibinfo  {journal} {Phys. Rev. Res.}\ }\textbf
  {\bibinfo {volume} {2}},\ \bibinfo {pages} {033293} (\bibinfo {year}
  {2020}{\natexlab{b}})}\BibitemShut {NoStop}%
\bibitem [{\citenamefont {Li}\ and\ \citenamefont {Ran}(2022)}]{li2022non}%
  \BibitemOpen
  \bibfield  {author} {\bibinfo {author} {\bibfnamefont {W.-M.}\ \bibnamefont
  {Li}}\ and\ \bibinfo {author} {\bibfnamefont {S.-J.}\ \bibnamefont {Ran}},\
  }\bibfield  {title} {\bibinfo {title} {Non-parametric semi-supervised
  learning in many-body hilbert space with rescaled logarithmic fidelity},\
  }\bibfield  {journal} {\bibinfo  {journal} {Mathematics}\ }\textbf {\bibinfo
  {volume} {10}},\ \href {https://doi.org/10.3390/math10060940}
  {10.3390/math10060940} (\bibinfo {year} {2022})\BibitemShut {NoStop}%
\bibitem [{\citenamefont {Kaneko}\ \emph {et~al.}(1996)\citenamefont {Kaneko},
  \citenamefont {Eguchi}, \citenamefont {Ohmatsu}, \citenamefont {Kakinuma},
  \citenamefont {Naruke}, \citenamefont {Suemasu},\ and\ \citenamefont
  {Moriyama}}]{kaneko1996peripheral}%
  \BibitemOpen
  \bibfield  {author} {\bibinfo {author} {\bibfnamefont {M.}~\bibnamefont
  {Kaneko}}, \bibinfo {author} {\bibfnamefont {K.}~\bibnamefont {Eguchi}},
  \bibinfo {author} {\bibfnamefont {H.}~\bibnamefont {Ohmatsu}}, \bibinfo
  {author} {\bibfnamefont {R.}~\bibnamefont {Kakinuma}}, \bibinfo {author}
  {\bibfnamefont {T.}~\bibnamefont {Naruke}}, \bibinfo {author} {\bibfnamefont
  {K.}~\bibnamefont {Suemasu}},\ and\ \bibinfo {author} {\bibfnamefont
  {N.}~\bibnamefont {Moriyama}},\ }\bibfield  {title} {\bibinfo {title}
  {Peripheral lung cancer: screening and detection with low-dose spiral ct
  versus radiography.},\ }\href
  {https://doi.org/10.1148/radiology.201.3.8939234} {\bibfield  {journal}
  {\bibinfo  {journal} {Radiology}\ }\textbf {\bibinfo {volume} {201}},\
  \bibinfo {pages} {798} (\bibinfo {year} {1996})},\ \bibinfo {note} {pMID:
  8939234}\BibitemShut {NoStop}%
\bibitem [{\citenamefont {Gordon}\ \emph {et~al.}(1985)\citenamefont {Gordon},
  \citenamefont {Szidon}, \citenamefont {Krotoszynski}, \citenamefont
  {Gibbons},\ and\ \citenamefont {O'Neill}}]{gordon1985volatile}%
  \BibitemOpen
  \bibfield  {author} {\bibinfo {author} {\bibfnamefont {S.~M.}\ \bibnamefont
  {Gordon}}, \bibinfo {author} {\bibfnamefont {J.~P.}\ \bibnamefont {Szidon}},
  \bibinfo {author} {\bibfnamefont {B.~K.}\ \bibnamefont {Krotoszynski}},
  \bibinfo {author} {\bibfnamefont {R.~D.}\ \bibnamefont {Gibbons}},\ and\
  \bibinfo {author} {\bibfnamefont {H.~J.}\ \bibnamefont {O'Neill}},\
  }\bibfield  {title} {\bibinfo {title} {{Volatile organic compounds in exhaled
  air from patients with lung cancer.}},\ }\href
  {https://doi.org/10.1093/clinchem/31.8.1278} {\bibfield  {journal} {\bibinfo
  {journal} {Clinical Chemistry}\ }\textbf {\bibinfo {volume} {31}},\ \bibinfo
  {pages} {1278} (\bibinfo {year} {1985})}\BibitemShut {NoStop}%
\bibitem [{\citenamefont {Yin}\ \emph {et~al.}(2021)\citenamefont {Yin},
  \citenamefont {Li}, \citenamefont {Lu}, \citenamefont {Yin}, \citenamefont
  {Su}, \citenamefont {Zeng}, \citenamefont {Luo}, \citenamefont {Ma},
  \citenamefont {Zhou}, \citenamefont {Orlandini}, \citenamefont {Yao},
  \citenamefont {Liu},\ and\ \citenamefont {Lang}}]{yin2021efficient}%
  \BibitemOpen
  \bibfield  {author} {\bibinfo {author} {\bibfnamefont {G.}~\bibnamefont
  {Yin}}, \bibinfo {author} {\bibfnamefont {L.}~\bibnamefont {Li}}, \bibinfo
  {author} {\bibfnamefont {S.}~\bibnamefont {Lu}}, \bibinfo {author}
  {\bibfnamefont {Y.}~\bibnamefont {Yin}}, \bibinfo {author} {\bibfnamefont
  {Y.}~\bibnamefont {Su}}, \bibinfo {author} {\bibfnamefont {Y.}~\bibnamefont
  {Zeng}}, \bibinfo {author} {\bibfnamefont {M.}~\bibnamefont {Luo}}, \bibinfo
  {author} {\bibfnamefont {M.}~\bibnamefont {Ma}}, \bibinfo {author}
  {\bibfnamefont {H.}~\bibnamefont {Zhou}}, \bibinfo {author} {\bibfnamefont
  {L.}~\bibnamefont {Orlandini}}, \bibinfo {author} {\bibfnamefont
  {D.}~\bibnamefont {Yao}}, \bibinfo {author} {\bibfnamefont {G.}~\bibnamefont
  {Liu}},\ and\ \bibinfo {author} {\bibfnamefont {J.}~\bibnamefont {Lang}},\
  }\bibfield  {title} {\bibinfo {title} {An efficient primary screening of
  covid-19 by serum raman spectroscopy},\ }\href
  {https://doi.org/https://doi.org/10.1002/jrs.6080} {\bibfield  {journal}
  {\bibinfo  {journal} {Journal of Raman Spectroscopy}\ }\textbf {\bibinfo
  {volume} {52}},\ \bibinfo {pages} {949} (\bibinfo {year} {2021})}\BibitemShut
  {NoStop}%
\bibitem [{\citenamefont {Smith}\ \emph {et~al.}(2013)\citenamefont {Smith},
  \citenamefont {Huser},\ and\ \citenamefont
  {Wachsmann-Hogiu}}]{smith2013raman}%
  \BibitemOpen
  \bibfield  {author} {\bibinfo {author} {\bibfnamefont {Z.~J.}\ \bibnamefont
  {Smith}}, \bibinfo {author} {\bibfnamefont {T.~R.}\ \bibnamefont {Huser}},\
  and\ \bibinfo {author} {\bibfnamefont {S.}~\bibnamefont {Wachsmann-Hogiu}},\
  }\bibfield  {title} {\bibinfo {title} {Raman scattering in pathology},\
  }\href@noop {} {\bibfield  {journal} {\bibinfo  {journal} {Biophotonics in
  Pathology}\ ,\ \bibinfo {pages} {207}} (\bibinfo {year} {2013})}\BibitemShut
  {NoStop}%
\bibitem [{\citenamefont {Auner}\ \emph {et~al.}(2018)\citenamefont {Auner},
  \citenamefont {Koya}, \citenamefont {Huang}, \citenamefont {Broadbent},
  \citenamefont {Trexler}, \citenamefont {Auner}, \citenamefont {Elias},
  \citenamefont {Mehne},\ and\ \citenamefont
  {Brusatori}}]{auner2018applications}%
  \BibitemOpen
  \bibfield  {author} {\bibinfo {author} {\bibfnamefont {G.~W.}\ \bibnamefont
  {Auner}}, \bibinfo {author} {\bibfnamefont {S.~K.}\ \bibnamefont {Koya}},
  \bibinfo {author} {\bibfnamefont {C.}~\bibnamefont {Huang}}, \bibinfo
  {author} {\bibfnamefont {B.}~\bibnamefont {Broadbent}}, \bibinfo {author}
  {\bibfnamefont {M.}~\bibnamefont {Trexler}}, \bibinfo {author} {\bibfnamefont
  {Z.}~\bibnamefont {Auner}}, \bibinfo {author} {\bibfnamefont
  {A.}~\bibnamefont {Elias}}, \bibinfo {author} {\bibfnamefont {K.~C.}\
  \bibnamefont {Mehne}},\ and\ \bibinfo {author} {\bibfnamefont {M.~A.}\
  \bibnamefont {Brusatori}},\ }\bibfield  {title} {\bibinfo {title}
  {Applications of raman spectroscopy in cancer diagnosis},\ }\href
  {https://doi.org/10.1007/s10555-018-9770-9} {\bibfield  {journal} {\bibinfo
  {journal} {Cancer and Metastasis Reviews}\ }\textbf {\bibinfo {volume}
  {37}},\ \bibinfo {pages} {691} (\bibinfo {year} {2018})}\BibitemShut
  {NoStop}%
\bibitem [{\citenamefont {Butler}\ \emph {et~al.}(2016)\citenamefont {Butler},
  \citenamefont {Ashton}, \citenamefont {Bird}, \citenamefont {Cinque},
  \citenamefont {Curtis}, \citenamefont {Dorney}, \citenamefont
  {Esmonde-White}, \citenamefont {Fullwood}, \citenamefont {Gardner},
  \citenamefont {Martin-Hirsch}, \citenamefont {Walsh}, \citenamefont
  {McAinsh}, \citenamefont {Stone},\ and\ \citenamefont
  {Martin}}]{butler2016using}%
  \BibitemOpen
  \bibfield  {author} {\bibinfo {author} {\bibfnamefont {H.~J.}\ \bibnamefont
  {Butler}}, \bibinfo {author} {\bibfnamefont {L.}~\bibnamefont {Ashton}},
  \bibinfo {author} {\bibfnamefont {B.}~\bibnamefont {Bird}}, \bibinfo {author}
  {\bibfnamefont {G.}~\bibnamefont {Cinque}}, \bibinfo {author} {\bibfnamefont
  {K.}~\bibnamefont {Curtis}}, \bibinfo {author} {\bibfnamefont
  {J.}~\bibnamefont {Dorney}}, \bibinfo {author} {\bibfnamefont
  {K.}~\bibnamefont {Esmonde-White}}, \bibinfo {author} {\bibfnamefont {N.~J.}\
  \bibnamefont {Fullwood}}, \bibinfo {author} {\bibfnamefont {B.}~\bibnamefont
  {Gardner}}, \bibinfo {author} {\bibfnamefont {P.~L.}\ \bibnamefont
  {Martin-Hirsch}}, \bibinfo {author} {\bibfnamefont {M.~J.}\ \bibnamefont
  {Walsh}}, \bibinfo {author} {\bibfnamefont {M.~R.}\ \bibnamefont {McAinsh}},
  \bibinfo {author} {\bibfnamefont {N.}~\bibnamefont {Stone}},\ and\ \bibinfo
  {author} {\bibfnamefont {F.~L.}\ \bibnamefont {Martin}},\ }\bibfield  {title}
  {\bibinfo {title} {Using raman spectroscopy to characterize biological
  materials},\ }\href {https://doi.org/10.1038/nprot.2016.036} {\bibfield
  {journal} {\bibinfo  {journal} {Nature Protocols}\ }\textbf {\bibinfo
  {volume} {11}},\ \bibinfo {pages} {664} (\bibinfo {year} {2016})}\BibitemShut
  {NoStop}%
\bibitem [{\citenamefont {Schie}\ and\ \citenamefont
  {Huser}(2013)}]{schie2013label}%
  \BibitemOpen
  \bibfield  {author} {\bibinfo {author} {\bibfnamefont {I.~W.}\ \bibnamefont
  {Schie}}\ and\ \bibinfo {author} {\bibfnamefont {T.}~\bibnamefont {Huser}},\
  }\bibfield  {title} {\bibinfo {title} {Label-free analysis of cellular
  biochemistry by raman spectroscopy and microscopy},\ }\href
  {https://doi.org/10.1002/cphy.c120025} {\bibfield  {journal} {\bibinfo
  {journal} {Comprehensive Physiology}\ }\textbf {\bibinfo {volume} {3}},\
  \bibinfo {pages} {941} (\bibinfo {year} {2013})}\BibitemShut {NoStop}%
\bibitem [{\citenamefont {Huang}\ \emph {et~al.}(2023)\citenamefont {Huang},
  \citenamefont {Sun}, \citenamefont {Sun}, \citenamefont {Shi}, \citenamefont
  {Chen}, \citenamefont {Ren}, \citenamefont {Ge}, \citenamefont {Jiang},
  \citenamefont {Liu}, \citenamefont {Knoll}, \citenamefont {Zhang},\ and\
  \citenamefont {Wang}}]{Huang2023}%
  \BibitemOpen
  \bibfield  {author} {\bibinfo {author} {\bibfnamefont {L.}~\bibnamefont
  {Huang}}, \bibinfo {author} {\bibfnamefont {H.}~\bibnamefont {Sun}}, \bibinfo
  {author} {\bibfnamefont {L.}~\bibnamefont {Sun}}, \bibinfo {author}
  {\bibfnamefont {K.}~\bibnamefont {Shi}}, \bibinfo {author} {\bibfnamefont
  {Y.}~\bibnamefont {Chen}}, \bibinfo {author} {\bibfnamefont {X.}~\bibnamefont
  {Ren}}, \bibinfo {author} {\bibfnamefont {Y.}~\bibnamefont {Ge}}, \bibinfo
  {author} {\bibfnamefont {D.}~\bibnamefont {Jiang}}, \bibinfo {author}
  {\bibfnamefont {X.}~\bibnamefont {Liu}}, \bibinfo {author} {\bibfnamefont
  {W.}~\bibnamefont {Knoll}}, \bibinfo {author} {\bibfnamefont
  {Q.}~\bibnamefont {Zhang}},\ and\ \bibinfo {author} {\bibfnamefont
  {Y.}~\bibnamefont {Wang}},\ }\bibfield  {title} {\bibinfo {title} {Rapid,
  label-free histopathological diagnosis of liver cancer based on raman
  spectroscopy and deep learning},\ }\href
  {https://doi.org/10.1038/s41467-022-35696-2} {\bibfield  {journal} {\bibinfo
  {journal} {Nature Communications}\ }\textbf {\bibinfo {volume} {14}},\
  \bibinfo {pages} {48} (\bibinfo {year} {2023})}\BibitemShut {NoStop}%
\bibitem [{\citenamefont {Maaten}\ and\ \citenamefont
  {Hinton}(2008)}]{van2008visualizing}%
  \BibitemOpen
  \bibfield  {author} {\bibinfo {author} {\bibfnamefont {L.~v.~d.}\
  \bibnamefont {Maaten}}\ and\ \bibinfo {author} {\bibfnamefont
  {G.}~\bibnamefont {Hinton}},\ }\bibfield  {title} {\bibinfo {title}
  {Visualizing data using t-sne},\ }\href
  {http://www.jmlr.org/papers/volume9/vandermaaten08a/vandermaaten08a.pdf}
  {\bibfield  {journal} {\bibinfo  {journal} {Journal of machine learning
  research}\ }\textbf {\bibinfo {volume} {9}},\ \bibinfo {pages} {2579}
  (\bibinfo {year} {2008})}\BibitemShut {NoStop}%
\bibitem [{\citenamefont {Yang}\ \emph {et~al.}(2021)\citenamefont {Yang},
  \citenamefont {Sun}, \citenamefont {Ran},\ and\ \citenamefont
  {Su}}]{yang2021visualizing}%
  \BibitemOpen
  \bibfield  {author} {\bibinfo {author} {\bibfnamefont {Y.}~\bibnamefont
  {Yang}}, \bibinfo {author} {\bibfnamefont {Z.-Z.}\ \bibnamefont {Sun}},
  \bibinfo {author} {\bibfnamefont {S.-J.}\ \bibnamefont {Ran}},\ and\ \bibinfo
  {author} {\bibfnamefont {G.}~\bibnamefont {Su}},\ }\bibfield  {title}
  {\bibinfo {title} {Visualizing quantum phases and identifying quantum phase
  transitions by nonlinear dimensional reduction},\ }\href
  {https://doi.org/10.1103/PhysRevB.103.075106} {\bibfield  {journal} {\bibinfo
   {journal} {Phys. Rev. B}\ }\textbf {\bibinfo {volume} {103}},\ \bibinfo
  {pages} {075106} (\bibinfo {year} {2021})}\BibitemShut {NoStop}%
\bibitem [{\citenamefont {Detterbeck}\ \emph {et~al.}(2016)\citenamefont
  {Detterbeck}, \citenamefont {Chansky}, \citenamefont {Groome}, \citenamefont
  {Bolejack}, \citenamefont {Crowley}, \citenamefont {Shemanski}, \citenamefont
  {Kennedy}, \citenamefont {Krasnik}, \citenamefont {Peake}, \citenamefont
  {Rami-Porta} \emph {et~al.}}]{detterbeck2016iaslc}%
  \BibitemOpen
  \bibfield  {author} {\bibinfo {author} {\bibfnamefont {F.~C.}\ \bibnamefont
  {Detterbeck}}, \bibinfo {author} {\bibfnamefont {K.}~\bibnamefont {Chansky}},
  \bibinfo {author} {\bibfnamefont {P.}~\bibnamefont {Groome}}, \bibinfo
  {author} {\bibfnamefont {V.}~\bibnamefont {Bolejack}}, \bibinfo {author}
  {\bibfnamefont {J.}~\bibnamefont {Crowley}}, \bibinfo {author} {\bibfnamefont
  {L.}~\bibnamefont {Shemanski}}, \bibinfo {author} {\bibfnamefont
  {C.}~\bibnamefont {Kennedy}}, \bibinfo {author} {\bibfnamefont
  {M.}~\bibnamefont {Krasnik}}, \bibinfo {author} {\bibfnamefont
  {M.}~\bibnamefont {Peake}}, \bibinfo {author} {\bibfnamefont
  {R.}~\bibnamefont {Rami-Porta}}, \emph {et~al.},\ }\bibfield  {title}
  {\bibinfo {title} {The iaslc lung cancer staging project: methodology and
  validation used in the development of proposals for revision of the stage
  classification of nsclc in the forthcoming (eighth) edition of the tnm
  classification of lung cancer},\ }\href
  {https://doi.org/https://doi.org/10.1016/j.jtho.2016.06.028} {\bibfield
  {journal} {\bibinfo  {journal} {Journal of thoracic oncology}\ }\textbf
  {\bibinfo {volume} {11}},\ \bibinfo {pages} {1433} (\bibinfo {year}
  {2016})}\BibitemShut {NoStop}%
\bibitem [{\citenamefont {Qiao}\ \emph {et~al.}(2018)\citenamefont {Qiao},
  \citenamefont {Su}, \citenamefont {Liu}, \citenamefont {Song}, \citenamefont
  {Luo}, \citenamefont {Mo},\ and\ \citenamefont {Wang}}]{qiao2018selective}%
  \BibitemOpen
  \bibfield  {author} {\bibinfo {author} {\bibfnamefont {X.}~\bibnamefont
  {Qiao}}, \bibinfo {author} {\bibfnamefont {B.}~\bibnamefont {Su}}, \bibinfo
  {author} {\bibfnamefont {C.}~\bibnamefont {Liu}}, \bibinfo {author}
  {\bibfnamefont {Q.}~\bibnamefont {Song}}, \bibinfo {author} {\bibfnamefont
  {D.}~\bibnamefont {Luo}}, \bibinfo {author} {\bibfnamefont {G.}~\bibnamefont
  {Mo}},\ and\ \bibinfo {author} {\bibfnamefont {T.}~\bibnamefont {Wang}},\
  }\bibfield  {title} {\bibinfo {title} {Selective surface enhanced raman
  scattering for quantitative detection of lung cancer biomarkers in
  superparticle@mof structure},\ }\href
  {https://doi.org/https://doi.org/10.1002/adma.201702275} {\bibfield
  {journal} {\bibinfo  {journal} {Advanced Materials}\ }\textbf {\bibinfo
  {volume} {30}},\ \bibinfo {pages} {1702275} (\bibinfo {year}
  {2018})}\BibitemShut {NoStop}%
\bibitem [{\citenamefont {Feng}\ \emph {et~al.}(2022)\citenamefont {Feng},
  \citenamefont {Miao}, \citenamefont {Zhang}, \citenamefont {Cui},
  \citenamefont {Wang}, \citenamefont {Feng}, \citenamefont {Gan},
  \citenamefont {Fu}, \citenamefont {Wang}, \citenamefont {Dai}, \citenamefont
  {Hu}, \citenamefont {Luo}, \citenamefont {Sun}, \citenamefont {Zhang},
  \citenamefont {Xiao}, \citenamefont {Wu}, \citenamefont {Zhou}, \citenamefont
  {Zou}, \citenamefont {He}, \citenamefont {Zhou},\ and\ \citenamefont
  {Han}}]{Feng2022}%
  \BibitemOpen
  \bibfield  {author} {\bibinfo {author} {\bibfnamefont {R.}~\bibnamefont
  {Feng}}, \bibinfo {author} {\bibfnamefont {Q.}~\bibnamefont {Miao}}, \bibinfo
  {author} {\bibfnamefont {X.}~\bibnamefont {Zhang}}, \bibinfo {author}
  {\bibfnamefont {P.}~\bibnamefont {Cui}}, \bibinfo {author} {\bibfnamefont
  {C.}~\bibnamefont {Wang}}, \bibinfo {author} {\bibfnamefont {Y.}~\bibnamefont
  {Feng}}, \bibinfo {author} {\bibfnamefont {L.}~\bibnamefont {Gan}}, \bibinfo
  {author} {\bibfnamefont {J.}~\bibnamefont {Fu}}, \bibinfo {author}
  {\bibfnamefont {S.}~\bibnamefont {Wang}}, \bibinfo {author} {\bibfnamefont
  {Z.}~\bibnamefont {Dai}}, \bibinfo {author} {\bibfnamefont {L.}~\bibnamefont
  {Hu}}, \bibinfo {author} {\bibfnamefont {Y.}~\bibnamefont {Luo}}, \bibinfo
  {author} {\bibfnamefont {W.}~\bibnamefont {Sun}}, \bibinfo {author}
  {\bibfnamefont {X.}~\bibnamefont {Zhang}}, \bibinfo {author} {\bibfnamefont
  {J.}~\bibnamefont {Xiao}}, \bibinfo {author} {\bibfnamefont {J.}~\bibnamefont
  {Wu}}, \bibinfo {author} {\bibfnamefont {B.}~\bibnamefont {Zhou}}, \bibinfo
  {author} {\bibfnamefont {M.}~\bibnamefont {Zou}}, \bibinfo {author}
  {\bibfnamefont {D.}~\bibnamefont {He}}, \bibinfo {author} {\bibfnamefont
  {X.}~\bibnamefont {Zhou}},\ and\ \bibinfo {author} {\bibfnamefont
  {X.}~\bibnamefont {Han}},\ }\bibfield  {title} {\bibinfo {title} {Single-atom
  sites on perovskite chips for record-high sensitivity and quantification in
  sers},\ }\href {https://doi.org/10.1007/s40843-022-1968-5} {\bibfield
  {journal} {\bibinfo  {journal} {Science China Materials}\ }\textbf {\bibinfo
  {volume} {65}},\ \bibinfo {pages} {1601} (\bibinfo {year}
  {2022})}\BibitemShut {NoStop}%
\bibitem [{\citenamefont {White}(1992)}]{W92DMRG}%
  \BibitemOpen
  \bibfield  {author} {\bibinfo {author} {\bibfnamefont {S.~R.}\ \bibnamefont
  {White}},\ }\bibfield  {title} {\bibinfo {title} {Density matrix formulation
  for quantum renormalization groups},\ }\href
  {https://doi.org/10.1103/PhysRevLett.69.2863} {\bibfield  {journal} {\bibinfo
   {journal} {Phys. Rev. Lett.}\ }\textbf {\bibinfo {volume} {69}},\ \bibinfo
  {pages} {2863} (\bibinfo {year} {1992})}\BibitemShut {NoStop}%
\bibitem [{\citenamefont {Sainath}\ \emph {et~al.}(2015)\citenamefont
  {Sainath}, \citenamefont {Vinyals}, \citenamefont {Senior},\ and\
  \citenamefont {Sak}}]{sainath2015convolutional}%
  \BibitemOpen
  \bibfield  {author} {\bibinfo {author} {\bibfnamefont {T.~N.}\ \bibnamefont
  {Sainath}}, \bibinfo {author} {\bibfnamefont {O.}~\bibnamefont {Vinyals}},
  \bibinfo {author} {\bibfnamefont {A.}~\bibnamefont {Senior}},\ and\ \bibinfo
  {author} {\bibfnamefont {H.}~\bibnamefont {Sak}},\ }\bibfield  {title}
  {\bibinfo {title} {Convolutional, long short-term memory, fully connected
  deep neural networks},\ }in\ \href
  {https://doi.org/10.1109/ICASSP.2015.7178838} {\emph {\bibinfo {booktitle}
  {2015 IEEE International Conference on Acoustics, Speech and Signal
  Processing (ICASSP)}}}\ (\bibinfo {year} {2015})\ pp.\ \bibinfo {pages}
  {4580--4584}\BibitemShut {NoStop}%
\bibitem [{\citenamefont {Ran}(2020)}]{R20MPSencode}%
  \BibitemOpen
  \bibfield  {author} {\bibinfo {author} {\bibfnamefont {S.-J.}\ \bibnamefont
  {Ran}},\ }\bibfield  {title} {\bibinfo {title} {Encoding of matrix product
  states into quantum circuits of one- and two-qubit gates},\ }\href
  {https://doi.org/10.1103/PhysRevA.101.032310} {\bibfield  {journal} {\bibinfo
   {journal} {Phys. Rev. A}\ }\textbf {\bibinfo {volume} {101}},\ \bibinfo
  {pages} {032310} (\bibinfo {year} {2020})}\BibitemShut {NoStop}%
\end{thebibliography}

\end{document}